\documentclass{PoS}
\pdfoutput=1
\usepackage{rotating,booktabs,amsmath,amssymb, multicol,url}
\usepackage[utf8]{inputenc}
\usepackage[numbers,sort&compress]{natbib}
\usepackage{natbibspacing}
\usepackage{ulem}

\title{Review on Composite Higgs Models}

\ShortTitle{Review on Composite Higgs Models}
\author{\speaker{Oliver Witzel}\\
        Department of Physics, University of Colorado Boulder, Boulder, CO, USA\\
        E-mail: \email{Oliver.Witzel@colorado.edu}
}
\abstract{Composite Higgs Models explore the possibility that the Higgs boson is an excitation of a new strongly interacting sector giving rise to electro-weak symmetry breaking. After describing how this new sector can be embedded into the Standard Model of elementary particle physics meeting experimental constraints, I will review efforts by the community to explore the physics of the new strong interaction using methods of lattice field theory. Challenges in understanding the numerical results are discussed and an outlook is given on possible future directions allowing to confirm or reject the composite Higgs hypothesis.}

\FullConference{The 36th Annual International Symposium on Lattice Field Theory - LATTICE2018\\
		22-28 July, 2018\\
		Michigan State University, East Lansing, Michigan, USA.}

\begin{document}
\section{Introduction}

The discovery of the Higgs boson by the Atlas and CMS experiments at the Large Hadron Collider (LHC) in 2012 \cite{Aad:2012tfa,Chatrchyan:2012ufa} completed the Standard Model (SM) by experimentally confirming the predicted Higgs boson. By now we know its mass,  $M_{H^0}=125.18(16)$ GeV \cite{Tanabashi:2018oca}, very precisely and a spin 0 interpretation is preferred over spin 2, whereas spin 1 is excluded due to observing $H^0\to \gamma\gamma$ decays. Determining its charge and parity quantum numbers is difficult because even and odd eigenstates can mix. Also the decay width predicted by the SM is too small for confirmation at the LHC. The LHC can however improve the precision on the Higgs couplings to SM particles and by that place limits on or eventually even find indirect signals of {\it new physics}\/ i.e.~physics not described by the SM. So far all direct searches for new physics have failed. Neither supersymmetric particles nor other, heavier resonances have been seen by the experiments. Hence it remains to discover the origin of the electro-weak sector.

An attractive idea to explain the origin of the electro-weak sector and avoid introducing a scalar as a fundamental particle is to extend the SM by a new strongly coupled gauge-fermion system. In such composite Higgs models, the Higgs boson arises as a bound state of this new, strongly interacting sector and its mass and quantum numbers match experimental values when accounting for SM interactions/corrections. For composite Higgs models to be viable, this new strongly interacting sector must exhibit a mechanism leading to a large separation of scales in order to explain why a 125 GeV Higgs boson but no other states have been found so far.  Furthermore, a mechanism to generate masses for SM fermions and gauge bosons is required and predictions from the composite Higgs model need to be in agreement with other experimental constraints like the $S$ parameter in the electro-weak sector.

Our current understanding of gauge-fermion systems suggests that the system cannot be QCD-like (e.g.~the generation of quark masses would be problematic) but obviously has to be chirally broken in order to predict massive resonances in the chiral limit. Near-conformal gauge theories are therefore of particular interest. 

Schematically composite Higgs models can be introduced \cite{Vecci:2016edi} by starting from a Higgs-less and massless SM (${\cal L}_{SM_0}$). A new sector describing the strong dynamics (${\cal L}_{SD}$) as well as a term describing the interactions between the new strong dynamics sector and the Higgs- and massless SM (${\cal L}_{int}$) are then added. These three terms describe all states of the SM plus the Higgs boson and other resonances originating from the strong sector or interactions with the SM

\begin{align}
  {\cal L}_{UV} \to {\cal L}_{SD} + {\cal L}_{SM_0} + {\cal L}_{int}\to {\cal L}_{SM} + \ldots
  \label{Eq.compHiggs}
\end{align}

It further contains a description to give mass to the SM gauge and fermion fields, the latter e.g.~via four-fermion interactions or partial compositeness. By construction, this description does not explain the mass of the fermions of the new strong dynamics. In that sense, composite Higgs models are effective models to explain the Higgs boson and the electro-weak sector but will themselves arise from some other theory in the UV (${\cal L}_{UV}$).

Composite Higgs models feature two general scenarios for the Higgs boson to arise: In the case of a ``dilaton-like'' particle, the Higgs is a light iso-singlet scalar ($0^{++}$) and the scale of the new strong sector is set by equating the pseudoscalar decay constant $F_\pi$ with the SM vev, i.e.~$F_\pi \sim 246$ GeV. Ideally the new sector has two massless flavors which give rise to three Goldstone bosons. These Goldstone bosons are in turn ``eaten'' which results in the longitudinal components of the $W^\pm$ and $Z^0$ boson. Examples for models focusing at this scenario are: an SU(2) gauge theory with two fundamental flavors explored by Drach et al.~\cite{Hietanen:2014xca,Arthur:2016dir,Arthur:2016ozw,Drach:2018prv},  the two-flavor sextet model with SU(3) gauge group investigated by LatHC \cite{Fodor:2012ty,Fodor:2015zna,Fodor:2016wal,Fodor:2016pls,Fodor:2019vmw} and others \cite{Hansen:2017ejh,Hasenfratz:2015ssa}, as well as an SU(3) gauge theory with eight fundamental flavors studied by the LatKMI \cite{Aoki:2013xza,Aoki:2016wnc} and the LSD collaboration \cite{Appelquist:2016viq,Appelquist:2018yqe}.

In the alternative scenario, the Higgs is a pseudo Nambu Goldstone Boson (pNGB) which arises due to spontaneous breaking of flavor symmetry (similar to pions in QCD). The Higgs acquires mass from its interaction and a non-trivial vacuum alignment introduces the angle $\chi$ as additional, free parameter i.e.~the scale is set by $F_\pi = (\text{SM vev})/ \sin(\chi) > 246$ GeV. This scenario requires strong dynamics with more than three flavors. Investigated examples include a model based on the SU(4)/Sp(4) coset studied by Bennett et al.~\cite{Bennett:2017kga,Lee:2018prv}, the two representation model by Ferretti (TACoS collaboration) \cite{Ayyar:2017qdf,Ayyar:2018zuk,Ayyar:2018ppa,Ayyar:2018glg}, as well as mass-split models (Hasenfratz et al.~and LSD collaboration) \cite{Brower:2014dfa,Brower:2015owo,Hasenfratz:2016gut,Hasenfratz:2017lne,Witzel:2018gxm}.\\

Essential for all composite Higgs models is near-conformal dynamics to exhibit a large separation of scales to explain experimental observations. In the following Section I will therefore start by introducing near-conformal gauge theories and some of the methods used to establish their properties. In Section \ref{Sec.Higgs_0++} I will review recent results of composite Higgs models where the Higgs is a light $0^{++}$ scalar and discuss implications of these findings for effective field theories (EFT)  describing a near-conformal, strong sector. Subsequently I present results for models featuring the Higgs as pNGB in Sec.~\ref{Sec.Higgs_pNGB}. Further developments are briefly discussed in Sec.~\ref{Sec.further} before summarizing in Sec.~\ref{Sec.summary}.

\section{Near-conformal gauge theories}
\begin{figure}[tb]
  \centering
  \includegraphics[scale=0.8]{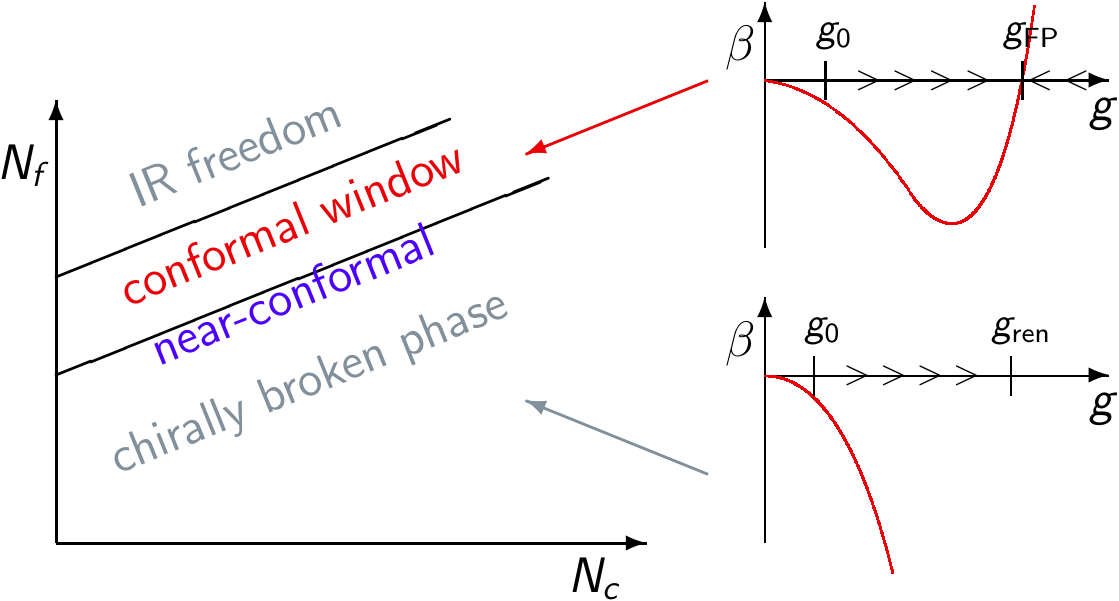}
  \caption{Sketch of the $N_f$-$N_c$ plane to visualize the range of near-conformal gauge theories and the characteristic shape of the $\beta$ function. }
  \label{fig.sketch_Nf_Nc}
\end{figure}

Gauge-fermion systems with $N_c$ colors and $N_f$ flavors in some representation are characterized by the renormalization group (RG) $\beta$ function which encodes how the gauge coupling $g^2$ changes w.r.t.~the energy scale. Assuming some fixed representation, we sketch the $N_f-N_c$ plane in Fig~\ref{fig.sketch_Nf_Nc}. Inspired by results derived from 2-loop perturbation theory, we expect that for a small number of flavors $N_f$ and $N_c\ge 2$ we find a chirally broken system which is basically QCD-like. The coupling runs fast and the $\beta$ function exhibits only the trivial, Gaussian fixed point at $g^2=0$. Keeping the number of colors fixed, but increasing the number of flavors, we expect that the $\beta$-function develops a second, infrared fixed point (IRFP) at $g^2>0$. For an even larger number of flavors, the system becomes IR free. The lowest number of flavors for which the $\beta$ function exhibits an IRFP defines the onset of the conformal window. In a conformal theory the gauge coupling takes the value at the fixed point and the masses of all bound states scale proportional to the fermion mass $m_f^{1/(1+\gamma^*)}$ with $\gamma^*$ the anomalous dimension. This property is named hyperscaling. In the zero mass limit, a length scale cannot be defined in conformal systems. Typically nonperturbative calculations are required to determine the onset of the conformal window\footnote{See Ref.~\cite{Dietrich:2006cm} for a characterization of different representations based on perturbative and Schwinger-Dyson arguments.}.

Near-conformal gauge theories live just below the onset of the conformal window and are thus chirally broken but inherit a slowly running (or walking coupling) due to the proximity of an IRFP.\\

Different nonperturbative methods can be used to establish the nature of a gauge-fermion system. Among others, scaling of hadron masses, the mode number of the Dirac operator, determinations of the anomalous dimension have been considered. In the case of the SU(2) gauge theory with fermions in the adjoint representation e.g., results of such investigations have lead to the following conclusions:
\begin{itemize}
  \item $N_f = 2$ is conformal \cite{Bergner:2016hip} 
  \item $N_f = 1$ likely conformal \cite{Athenodorou:2014eua}
  \item $N_f = 3/2$ (3 Majorana fermions) is conformal with anomalous dimension $\gamma^*=0.37(2)$ \cite{Bergner:2018fxm}
  \item $N_f = 1/2$ (1 Majorana fermion) is QCD-like \cite{Bergner:2015adz}
\end{itemize}

An alternative and potentially more powerful method is the nonperturbative determination of the $\beta$ function itself. If the $\beta$ function exhibits an IRFP (i.e.~a zero at $g^2>0$), the system is conformal. Numerically one calculates the discretized $\beta$ or step-scaling function which requires calculations on a set of different volumes to subsequently take the infinite volume continuum limit. Such calculations are well established in QCD \cite{Luscher:1991wu}  using the Schr\"odinger functional scheme. Modern calculations take advantage of the fact that the gradient flow defines a renormalized coupling \cite{Luscher:2010iy,Fodor:2012td,Fodor:2014cpa} 
\begin{align}
  g^2_c (L) &= \frac{128 \pi^2}{3(N_c^2 -1)}\;\frac{1}{C(c,L)} t^2 \langle E(t) \rangle
  \quad \text{with} \; \sqrt{8t} =c\cdot L.
  \label{eq.flow_gc2}
\end{align}
In Eq.~(\ref{eq.flow_gc2}), $E(t)$ denotes the energy density at gradient flow time $t$, $L$ is the extent of the lattice with volume $L^4$, and $c$ defines the scheme of the calculation. The factor $1/C(c,L)$ introduced in \cite{Fodor:2014cpa} corrects for finite volume and discretization artifacts. Calculating $g_c^2$ on volumes $L$ and $sL$ allows then to define a discrete $\beta$ function for a scale change $s$
\begin{align}
  \beta_s^{c}(g^2_c;L) &= \frac{g^2_c(sL) - g^2(L)}{\text{log}(s^2)}.
  \label{eq.flow_beta}
\end{align}

\begin{figure}[tb]
  \begin{picture}(148,145)
  \put(3,100){\includegraphics[height=0.2\textheight]{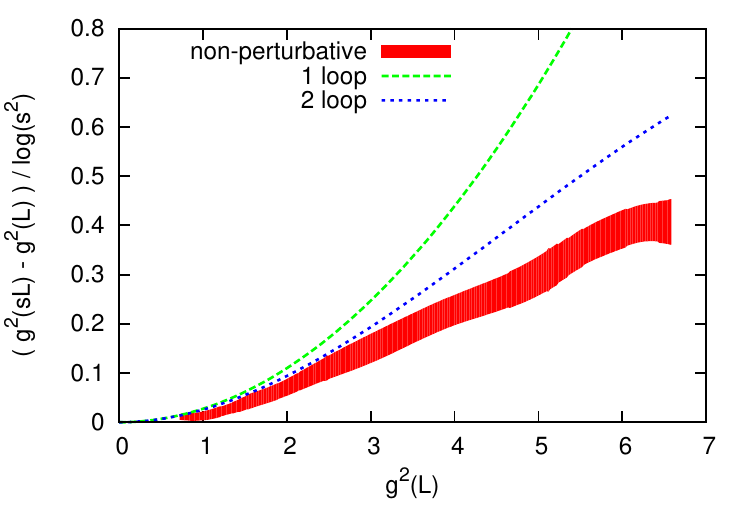}}
  \put(84,100){\includegraphics[height=0.2\textheight]{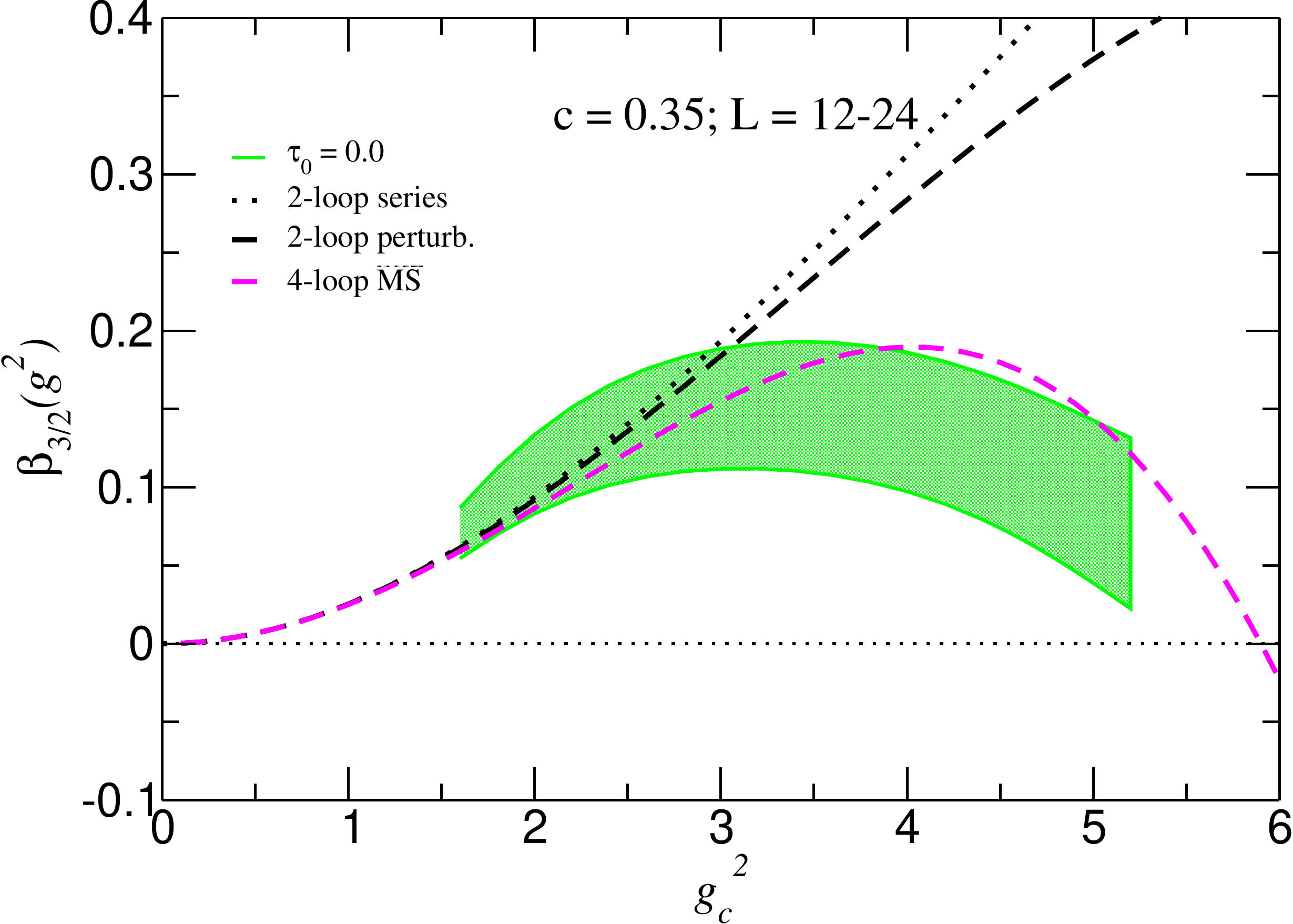}}
   \put(9,50){\includegraphics[height=0.2\textheight]{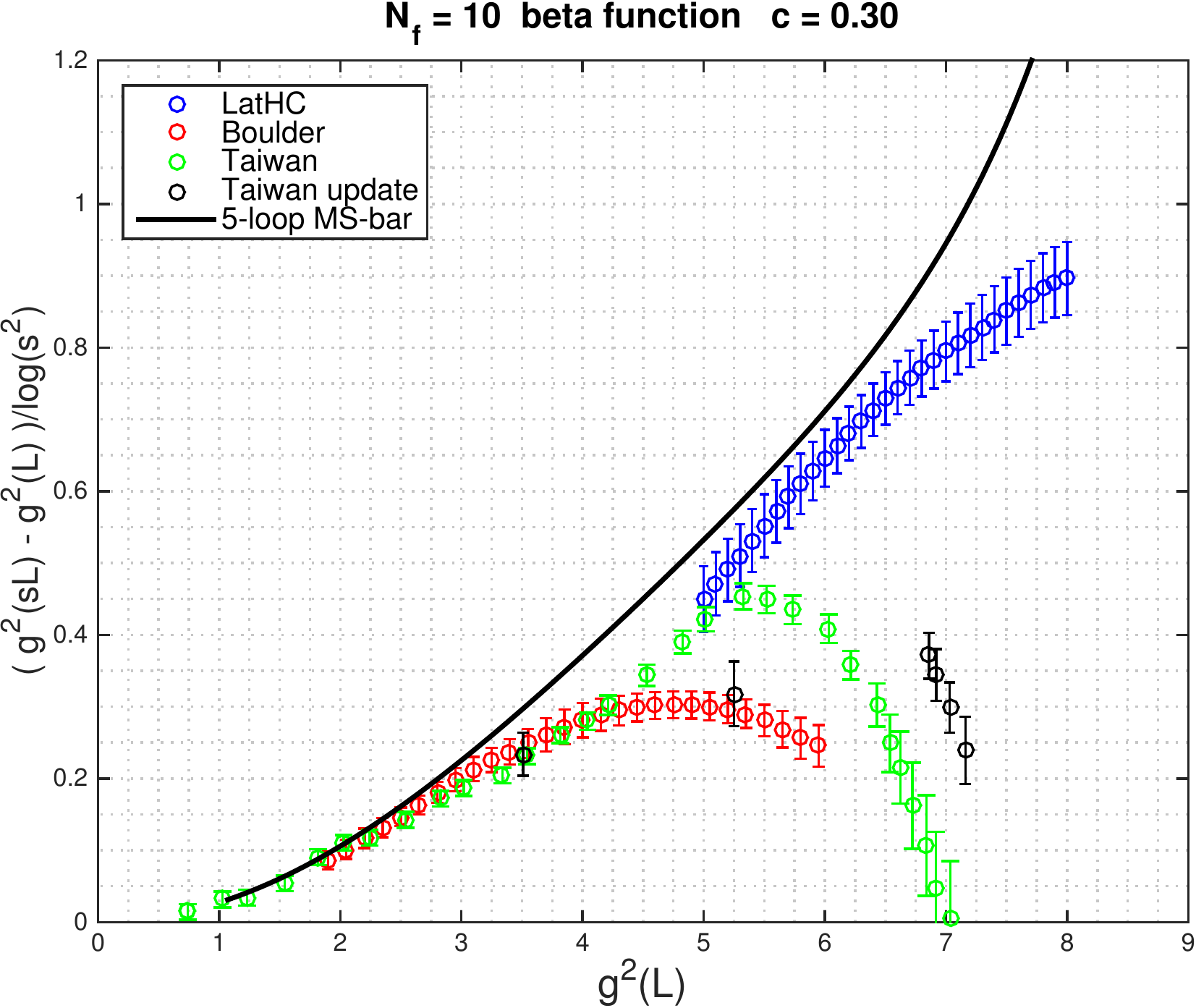}}
   \put(85,50){\includegraphics[height=0.2\textheight]{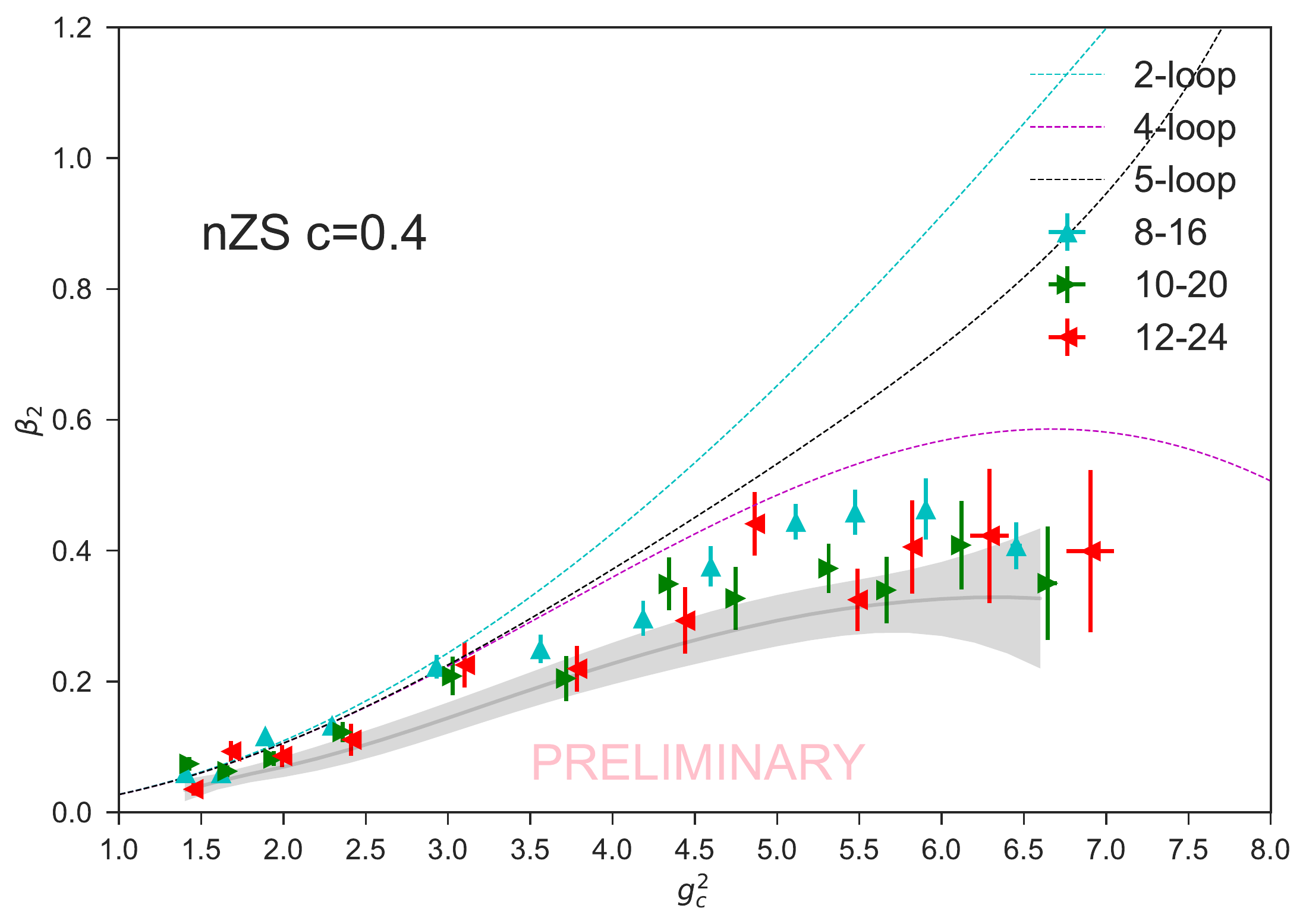}}
  \put(8,0){\includegraphics[height=0.2\textheight]{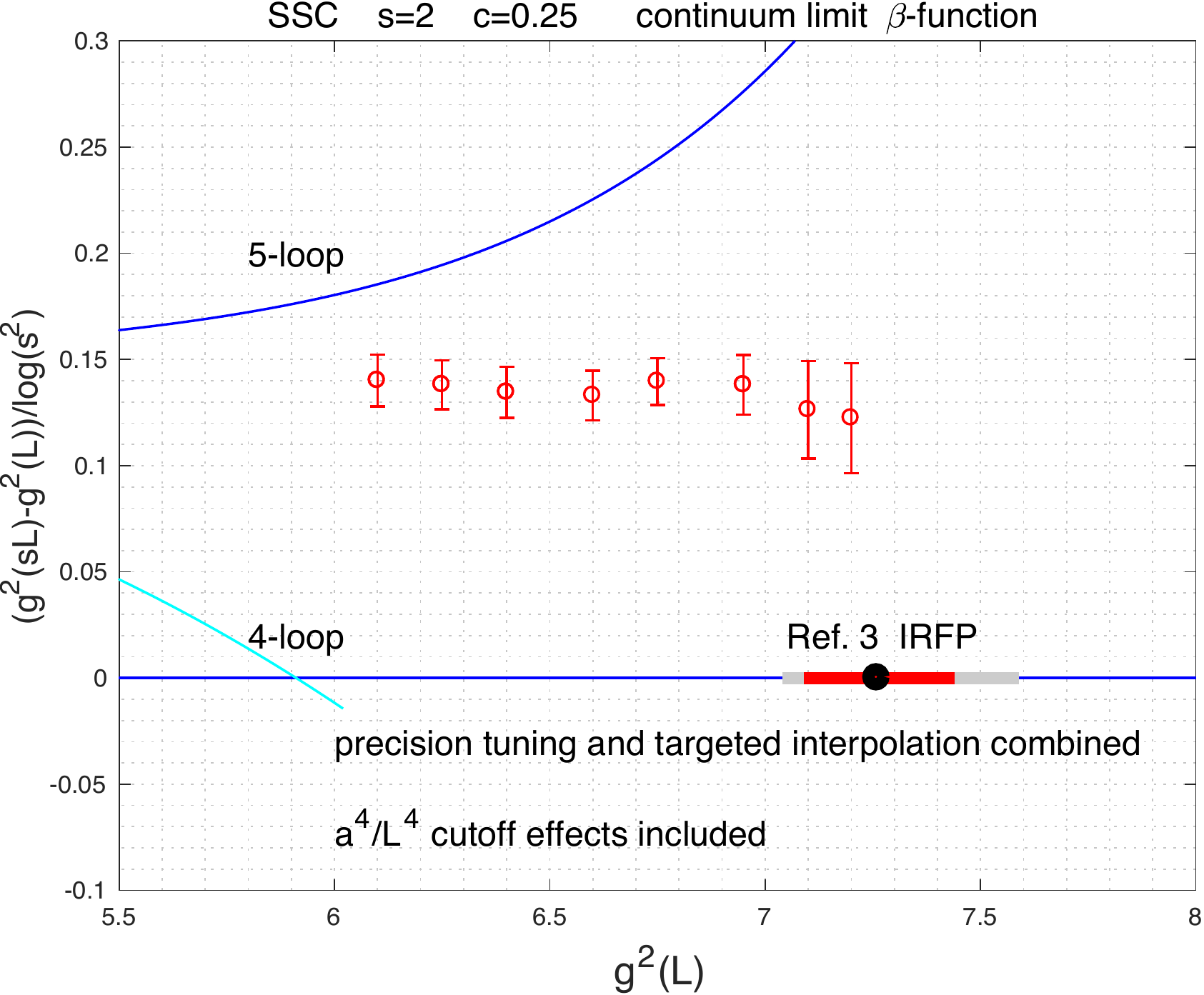}}
  \put(80,0){\includegraphics[height=0.2\textheight]{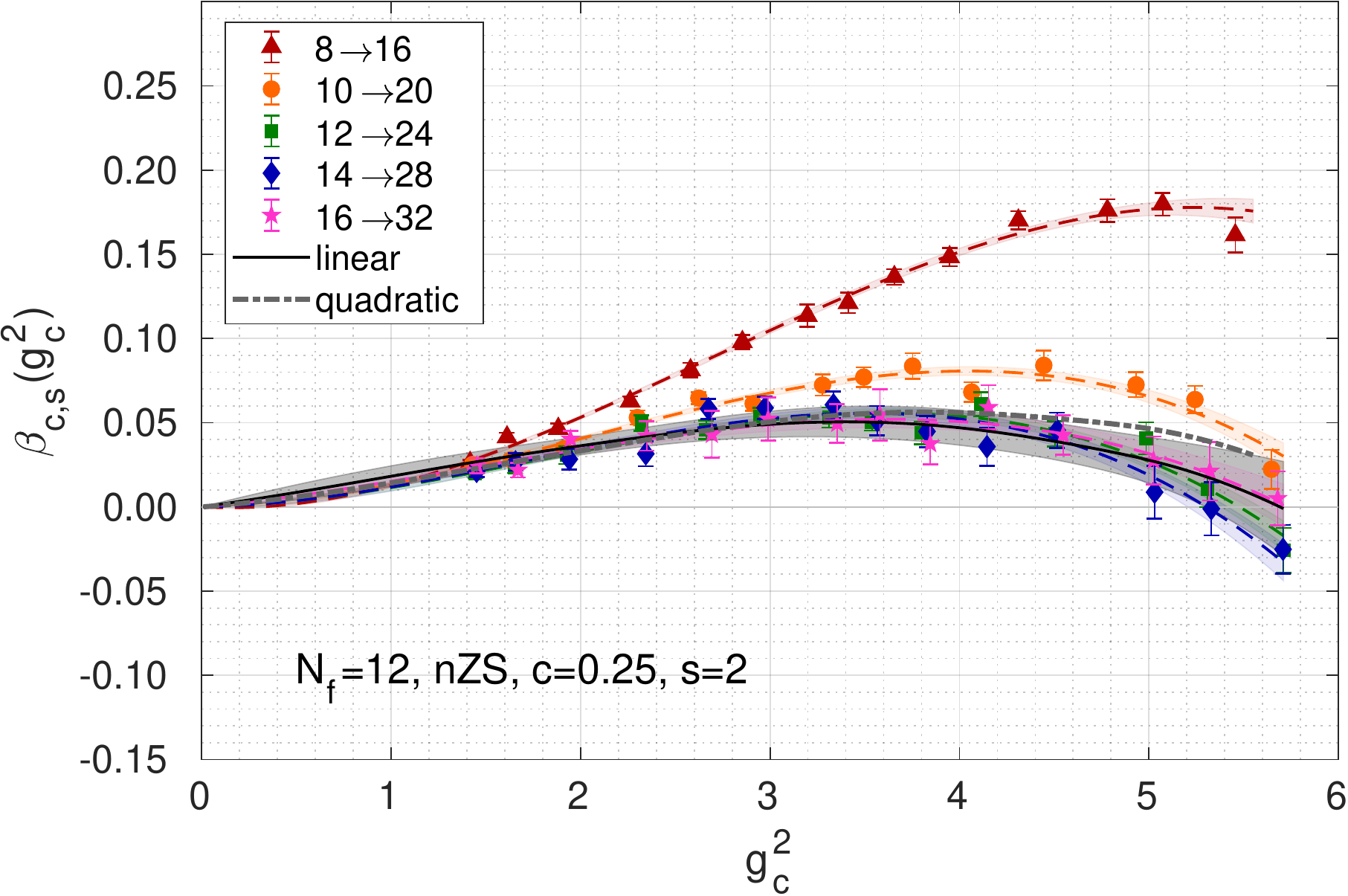}}
  \end{picture}
  \caption{Nonperturbatively determined step-scaling functions for SU(3) gauge theories with two sextet flavors (top row) from Refs.~\cite{Fodor:2015zna,Hasenfratz:2015ssa}, ten fundamental flavors (middle row) from Refs.~\cite{Fodor:2018tdg,Hasenfratz:2017qyr} also showing data from \cite{Chiu:2016uui,Chiu:2017kza,Chiu:2018edw}, and twelve fundamental flavors (bottom row) from Refs.~\cite{Fodor:2017gtj,Hasenfratz:2018wpq}. Plots in the left column highlight results obtained with staggered fermions; plots in the right column calculations performed with Wilson or domain wall fermions. }
  \label{fig.step-scaling}
\end{figure}

Repeating this calculation for a set of volume pairs, the $L\to \infty$ limit can be taken to remove discretization effects and obtain the continuum limit which can be compared to perturbative predictions calculated up to 5-loop order \cite{Baikov:2016tgj,Ryttov:2016ner}. The proper continuum limit result is free of discretization effects and hence results based on different actions, flows, operators, etc.~are expected to agree. In practice, however, agreement between different calculations is not always observed as can be seen by the plots in Fig.~\ref{fig.step-scaling} for different systems with SU(3) gauge group.\footnote{In addition new results for the non-controversial system SU(3) with 13 fundamental flavor were presented \cite{Fodor:2018uih}.}

Understanding the origin of these discrepancies is active research. Possible sources are:\\
\textbf{Lattice volumes:} results are only expected to agree in the continuum limit after performing the $L\to\infty$ extrapolation. Hence simulations at large enough volumes and with sufficiently many volume pairs are required in order to take a reliable continuum limit. While at weak coupling, perturbation theory provides guidance to the functional form of the extrapolation, corrections to that form may be significant at strong coupling. Further it is known that schemes with smaller $c$-values require simulations on larger volumes compared to schemes with larger $c$-values. Increasing $c$ however also triggers the statistical uncertainties to grow. \\
\textbf{Discretization effects:} different actions, flows, and operators have different discretization errors. Hence a certain volume for scheme $c$ may be sufficient for some combination but not for others. Interesting in this respect is that some combinations feature a fully $O(a^2)$ improved set-up \`a la Symanzik. An example is the Symanzik gauge action combined with Zeuthen flow \cite{Ramos:2015baa} and the Symanzik operator to determine the energy density. Further, the perturbative tree-level normalization \cite{Fodor:2014cpa} improves the results for domain wall fermions impressively well for $N_f$ = 12 and 10 over the entire range in $g_c^2$ studied, whereas perturbative improvement breaks down for $N_f$ = 8 staggered fermion simulations \cite{Fodor:2015baa}.\\
\textbf{Fermion action:} discrepancies are seen for results based on staggered fermions and either Wilson or domain wall calculation. Since universality of fermion formulations has so far only been investigated in QCD-like systems at the Gaussian fixed point (see e.g.~\cite{Sharpe:2006re}) that may give rise to the question: Are staggered, Wilson, and domain wall fermions in conformal systems investigating the same IRFP? \cite{Hasenfratz:2017mdh,Hasenfratz:2017qyr}
\section{Higgs as a light $0^{++}$ scalar}
\label{Sec.Higgs_0++}
In this Section we discuss three examples containing a light $0^{++}$ scalar as candidate for the Higgs boson. We start out by presenting the SU(2) gauge theory with two flavors in the fundamental representation before moving on to systems with SU(3) gauge group, two sextet flavors and eight fundamental flavors. 
\subsection{SU(2) gauge theory with two fundamental flavors}

\begin{figure}[tb]
   \includegraphics[width=0.47\textwidth]{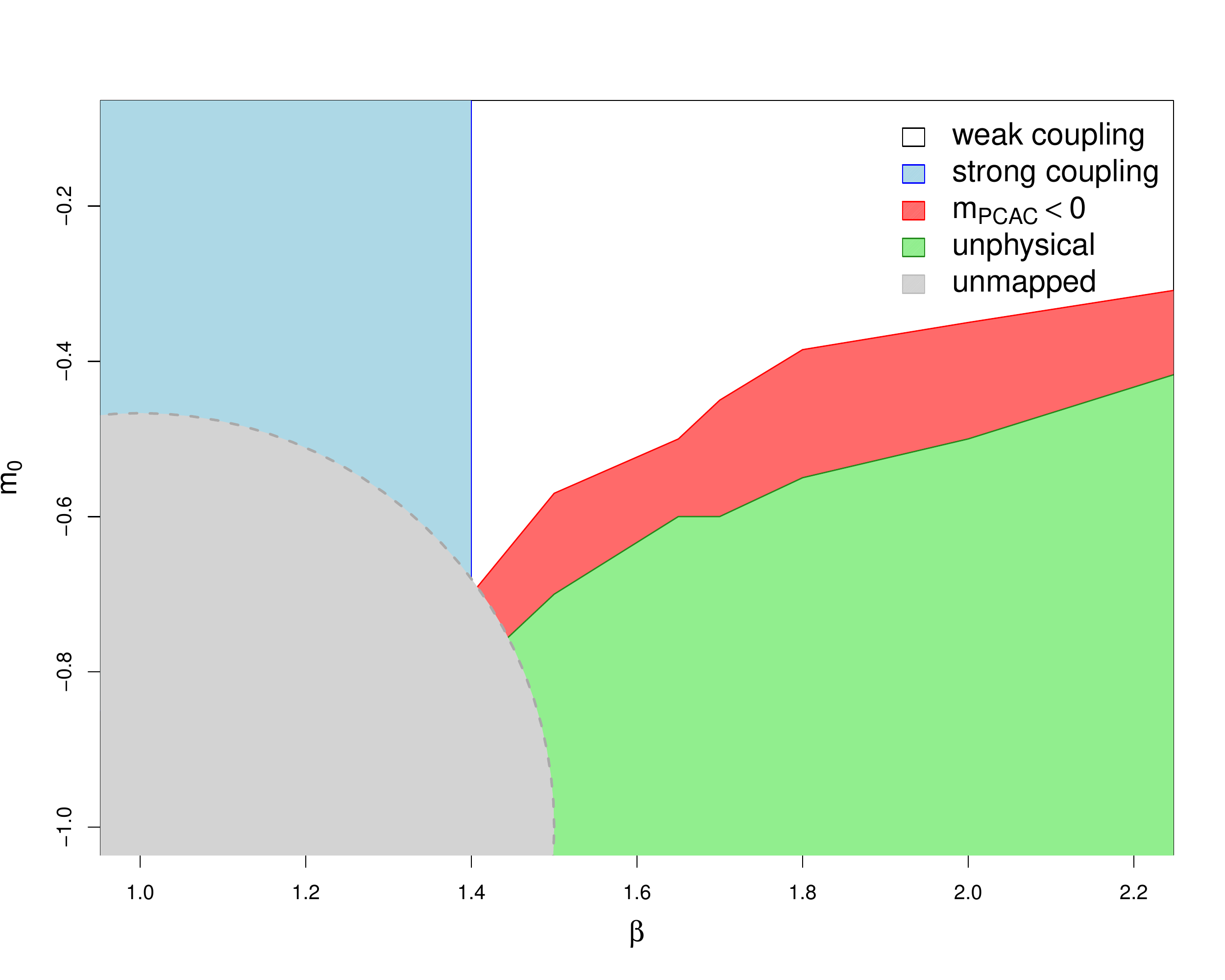} \hfill
   \includegraphics[width=0.47\textwidth]{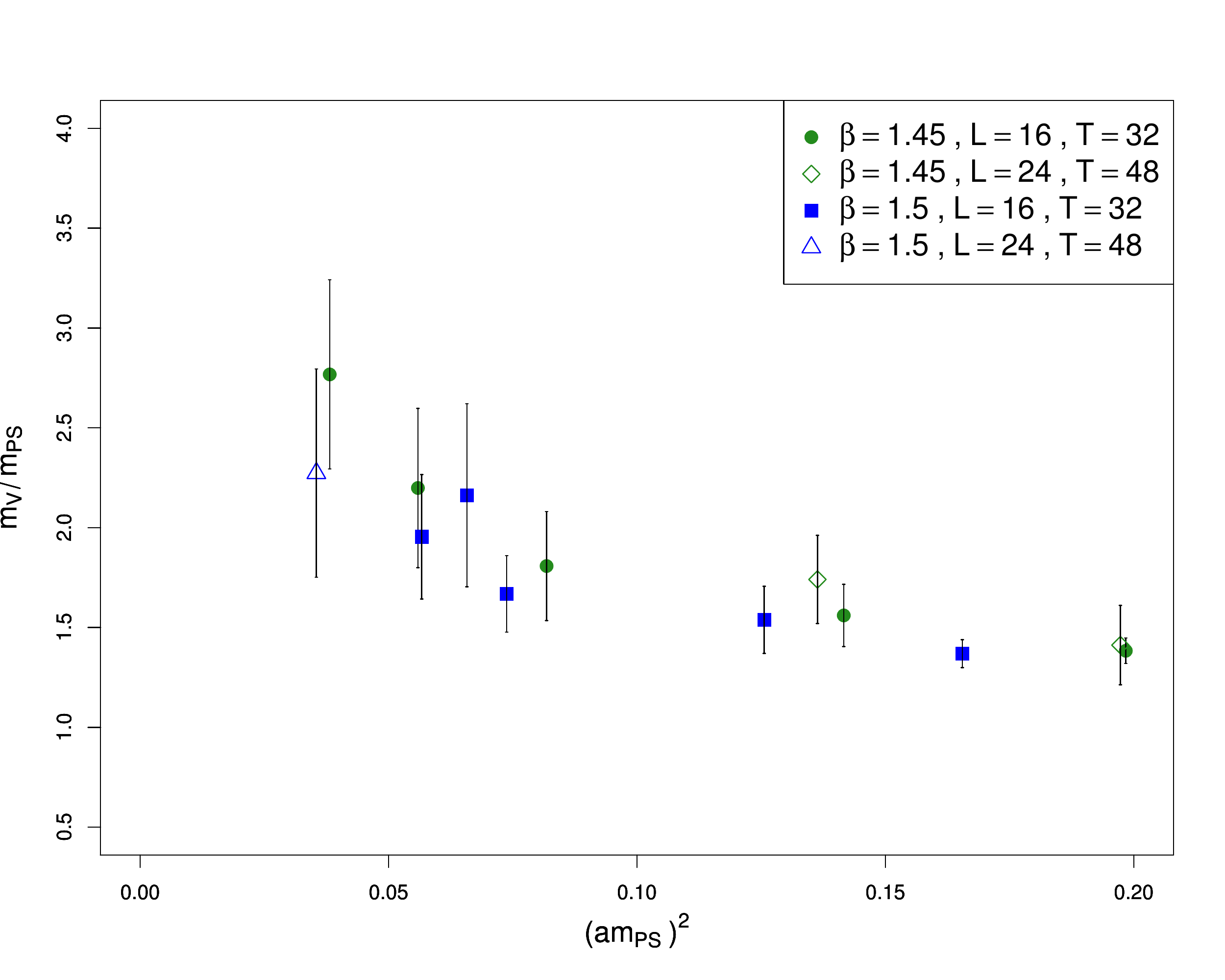}
   \caption{Left: current status of the exploration of the phase diagram using the improved action; right: chiral behavior of the ratio of $m_V/m_{PS}$ which for a chirally broken system is expected to diverge in the chiral limit. Plots courtesy by Drach \cite{Drach:2018prv}.}
   \label{fig.SU2.2f}
\end{figure}

A realization of a composite Higgs model with the minimal flavor content is given by an SU(2) gauge theory with two fundamental flavors. In order to improve investigations performed with unimproved Wilson fermions and plaquette gauge action \cite{Arthur:2016dir,Arthur:2016ozw}, Drach, Janowski, Pica, and Prelovsek have started a new project to generate dynamical gauge field configurations using Wilson-clover fermions and Symanzik gauge action. The left plot in Fig.~\ref{fig.SU2.2f} summarizes their current knowledge on the phase diagram which so far reveals little changes w.r.t.~the unimproved set-up. In order to identify the chiral regime, the ratio of vector over pseudoscalar meson mass, $m_V/m_{PS}$, is studied and presented in the right plot of Fig.~\ref{fig.SU2.2f}. Despite large uncertainties for lighter quark masses, a diverging behavior in the chiral limit is already emerging. Further insight into that will be revealed from a planned investigation of scattering processes and the determination to the $\rho\pi\pi$ coupling \cite{Drach:2018prv}.

\subsection{SU(3) with $N_f = 2$ sextet flavors (two-index symmetric representation)}
\begin{figure}[tb]
  \centering
  \includegraphics[height=0.26\textheight]{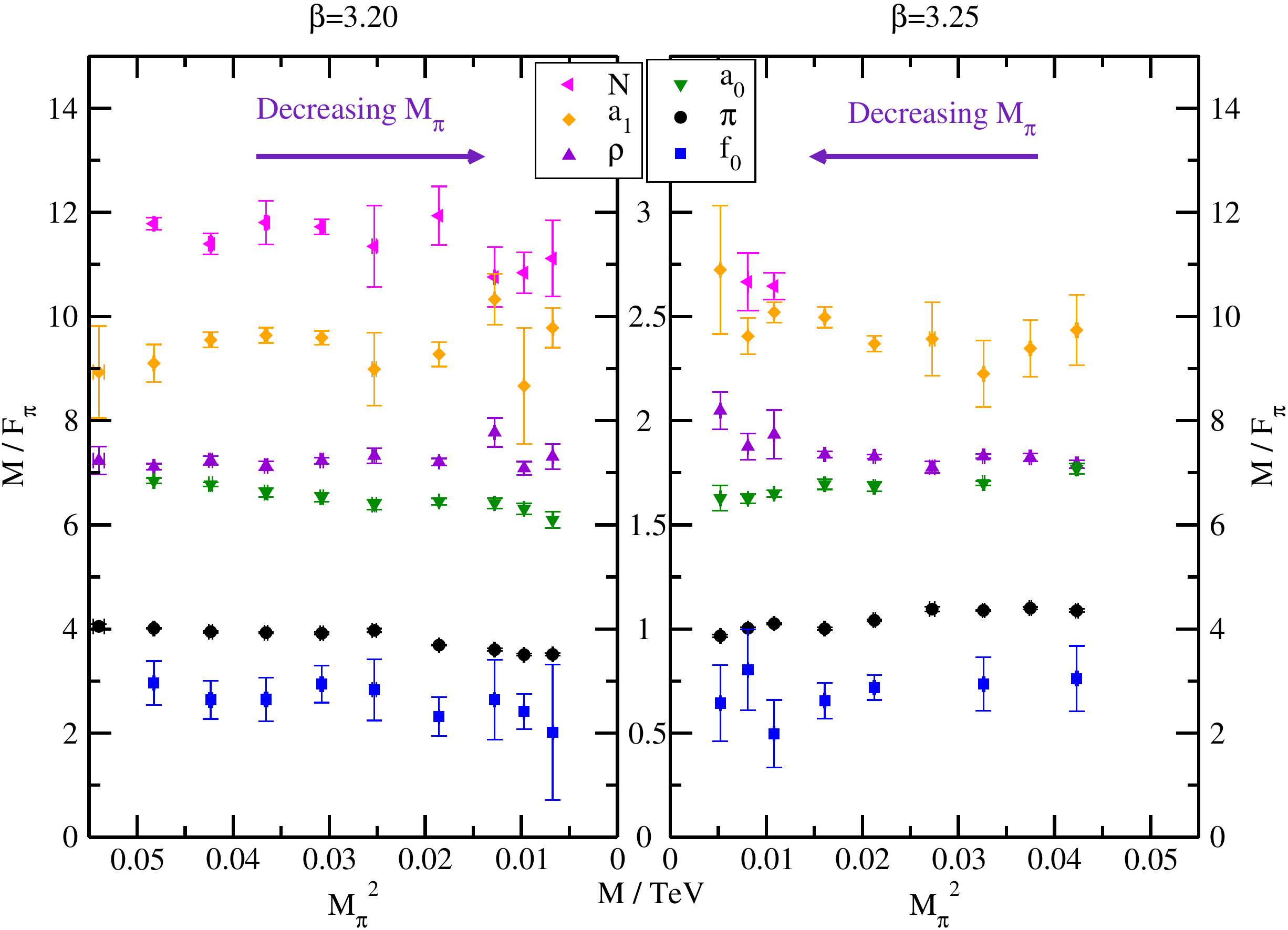}
  \caption{Spectrum in units of $F_\pi$ for the SU(3) sextet model with two flavors from Ref.~\cite{Fodor:2016pls}. Two values of the bare gauge coupling, $\beta=3.20$ (left panel) and $\beta=3.25$ (right panel), are investigated and simulations performed with rooted staggered fermions. The iso-singlet scalar ($f_0$) is the lightest state, close to the pseudoscalar ($\pi$) and separated from the multiplet scalar ($a_0$), vector ($\rho$), axial ($a_1$), and the nucleon ($N$).}
  \label{fig.SU3_sextet_LatHC} 
\end{figure}

The two flavor sextet model has attracted quite some attention because it likewise features the minimal flavor content to describe electro-weak symmetry breaking and is expected to be very close to the onset of the conformal window. The LatHC collaboration concludes from their results obtained with rooted staggered fermions that the system is chirally broken  which is also supported by their investigation of the step-scaling function up to $g^2 \sim 6.5$. Moreover, the spectrum exhibits a light $0^{++}$ scalar (labeled in analogy to QCD $f_0$) \cite{Fodor:2012ty}. In Fig.~\ref{fig.SU3_sextet_LatHC}, the results for the spectrum are presented as dimensionless ratios of masses over the pseudoscalar decay constants $F_\pi$ and are obtained at two different values of the bare gauge coupling, $\beta=3.20$ and 3.25. They exhibit little scale dependence and show a $0^{++}$ which is the lowest state, even slightly lighter than the pseudoscalar pions. In the chiral limit, a chirally broken theory predicts the pions to be massless, while the $0^{++}$ is expected to have finite mass. Details of an analysis based on the dilaton EFT are presented in Ref.~\cite{Fodor:2019vmw}.

\begin{figure}[tb]
  \centering
  \includegraphics[height=0.22\textheight]{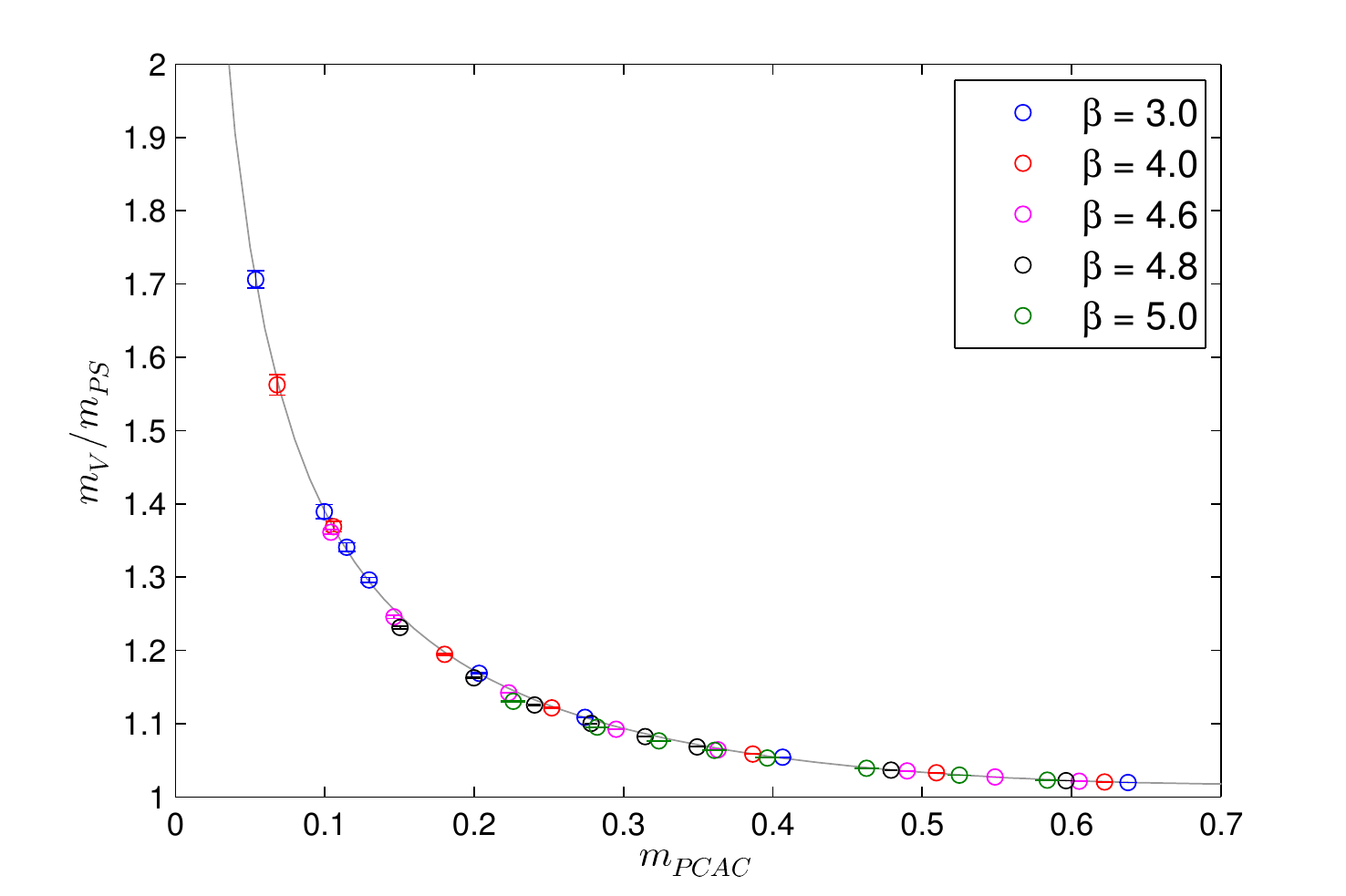}  \hfill
  \includegraphics[height=0.22\textheight]{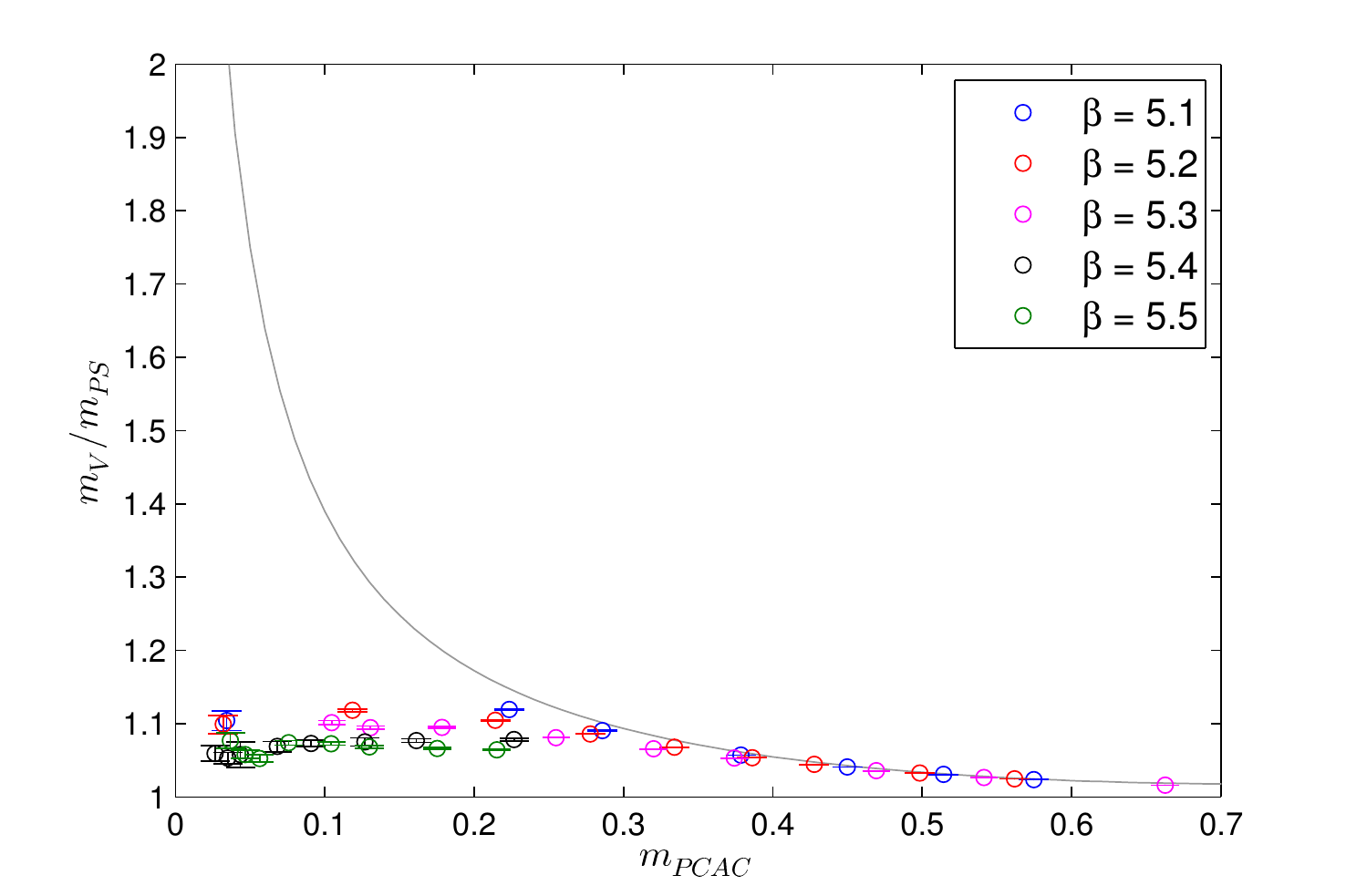}
  \caption{Ratio of vector over pseudoscalar mass, $m_V/m_{PS}$, as function of the PCAC mass from Ref.~\cite{Hansen:2017ejh}. These simulations of the two flavor SU(3) sextet model are performed with Wilson fermions and plaquette gauge action.  At strong couplings (left) a diverging, QCD-like behavior is observed which is interpreted as a chirally broken phase. At weak couplings (right) a flat behavior corresponding to hyperscaling in a conformal phase is seen.}
  \label{fig.SU3_sextet_CP3}
\end{figure}

The investigation of the same model by Hansen, Drach, and Pica is based on Wilson fermions and identified two phases depending on the strength of the gauge coupling: one seemingly chirally broken at stronger couplings; one looking IR conformal at weaker couplings \cite{Hansen:2017ejh}. Using the ratio of the vector over the pseudoscalar mass as an example, Fig.~\ref{fig.SU3_sextet_CP3} shows on the left a diverging ratio (similar to QCD-like systems), whereas the ratio on the right is flat as expected for a conformal system exhibiting hyperscaling. The latter is also in agreement with the indications of a possible IRFP observed in a determination of step-scaling function by  Hasenfratz, Liu, Yu-Han Huang \cite{Hasenfratz:2015ssa} using nHYP-smeared Wilson-clover fermions. 
 
\subsection{SU(3) with $N_f = 8$ fundamental flavors}
\begin{figure}[tb]
  \centering
  \includegraphics[height=0.22\textheight]{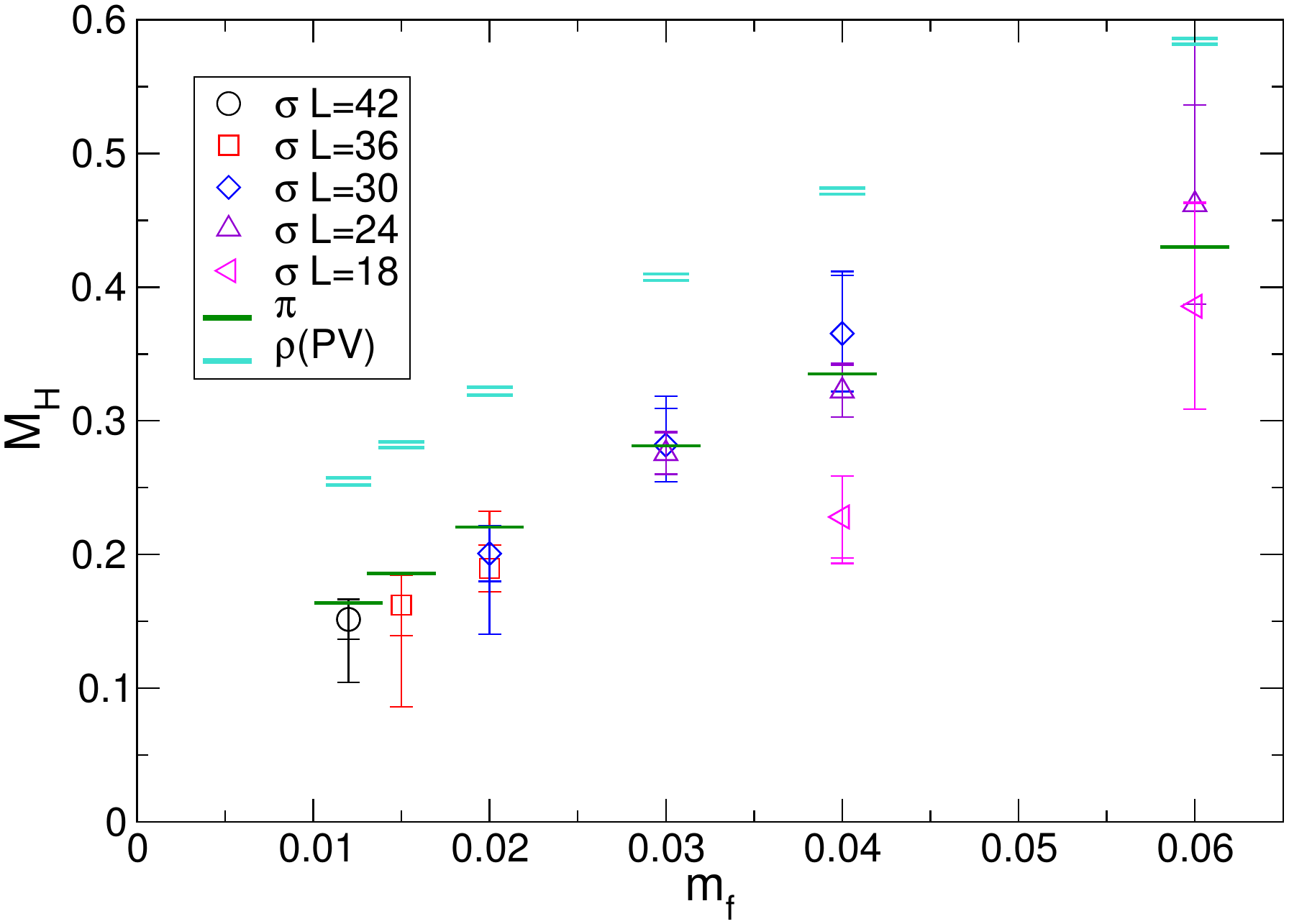} \hfill
  \includegraphics[height=0.22\textheight]{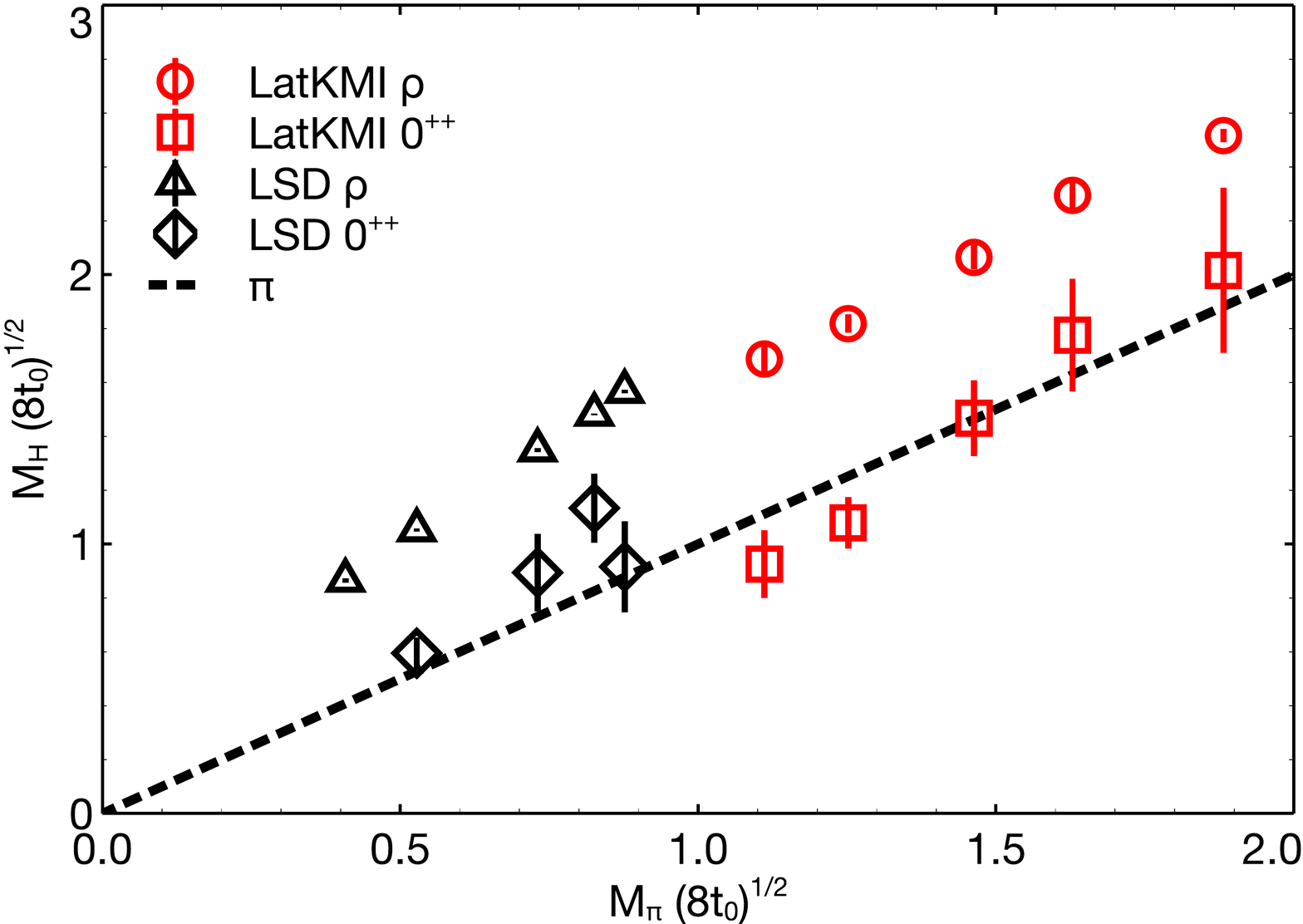}
  \caption{Observation of a light $0^{++}$ scalar ($\sigma$) in relation to the pseudoscalar ($\pi$) and vector ($\rho$) for an SU(3) gauge theory with eight fundamental flavors. The plot on the left from Ref.~\cite{Aoki:2016wnc} shows details of the results obtained by the LatKMI collaboration, whereas the plot on the right from Ref.~\cite{Appelquist:2016viq} compares results by LatKMI and LSD in units of $\sqrt{8t_0}$.}
  \label{fig.SU3_8f}
\end{figure}

The SU(3) gauge fermion system with eight fundamental flavors is mostly considered to be chirally broken but close to the onset to of the conformal window. Despite using different methods and enormous numerical efforts, no final conclusion has been reached. Investigations include a step-scaling analysis of the discrete  $\beta$ function \cite{Hasenfratz:2014rna,Fodor:2015baa}, explorations of the finite temperature phase diagram \cite{Deuzeman:2008sc,Jin:2010vm,Schaich:2012fr}, or studies of the low-lying meson spectrum \cite{Aoki:2014oha,Appelquist:2016viq,Aoki:2016wnc,Appelquist:2018yqe}. Since a theory with eight flavors exhibits 63 Goldstone bosons, it is not an ideal candidate to explain electro-weak symmetry breaking. While it is possible to reduce the number of light Goldstones by assigning e.g.~mass or charge to some flavors, also investigations of the degenerate eight flavor theory are worthwhile and allow to study features of near-conformal gauge theories. Of particular interest is the fact that two groups independently observe a light $0^{++}$ scalar, degenerate with the pion. As can be seen in the left plot of Fig.~\ref{fig.SU3_8f} showing data from the LatKMI collaboration, the $0^{++}$ is much lighter than the vector (rho) and degenerate with the pseudoscalar (pion). Although using different staggered fermion formulations, the LatKMI results are in good agreement with the results by the LSD collaboration. Turning both results into dimensionless ratios using the Wilson flow scale $\sqrt{8t_0}$, the right plot in Fig.~\ref{fig.SU3_8f} shows that LSD's data basically continue LatKMI's data to significantly smaller quark masses. This also holds for the $0^{++}$ which is an iso-singlet scalar and thus receives quark-line connected and disconnected contributions. Consequently the determination of this state is more difficult and numerically expensive. Disconnected contributions are calculated using stochastic estimators and in addition one has to account for a large vacuum subtraction because the $0^{++}$ has the same quantum numbers as the vacuum. Details of the calculation as well as alternative ideas how to improve the determination were presented by Rebbi \cite{Rebbi:2019fuk}.   Compared to QCD, the determination of the $0^{++}$ is however somewhat easier because the $0^{++}$ is lighter in near-conformal systems than in QCD. Thus depending on the masses simulated it can be a stable particle which is energetically protected from decaying.

\begin{figure}[tb]
  \centering  
  \includegraphics[height=0.21\textheight]{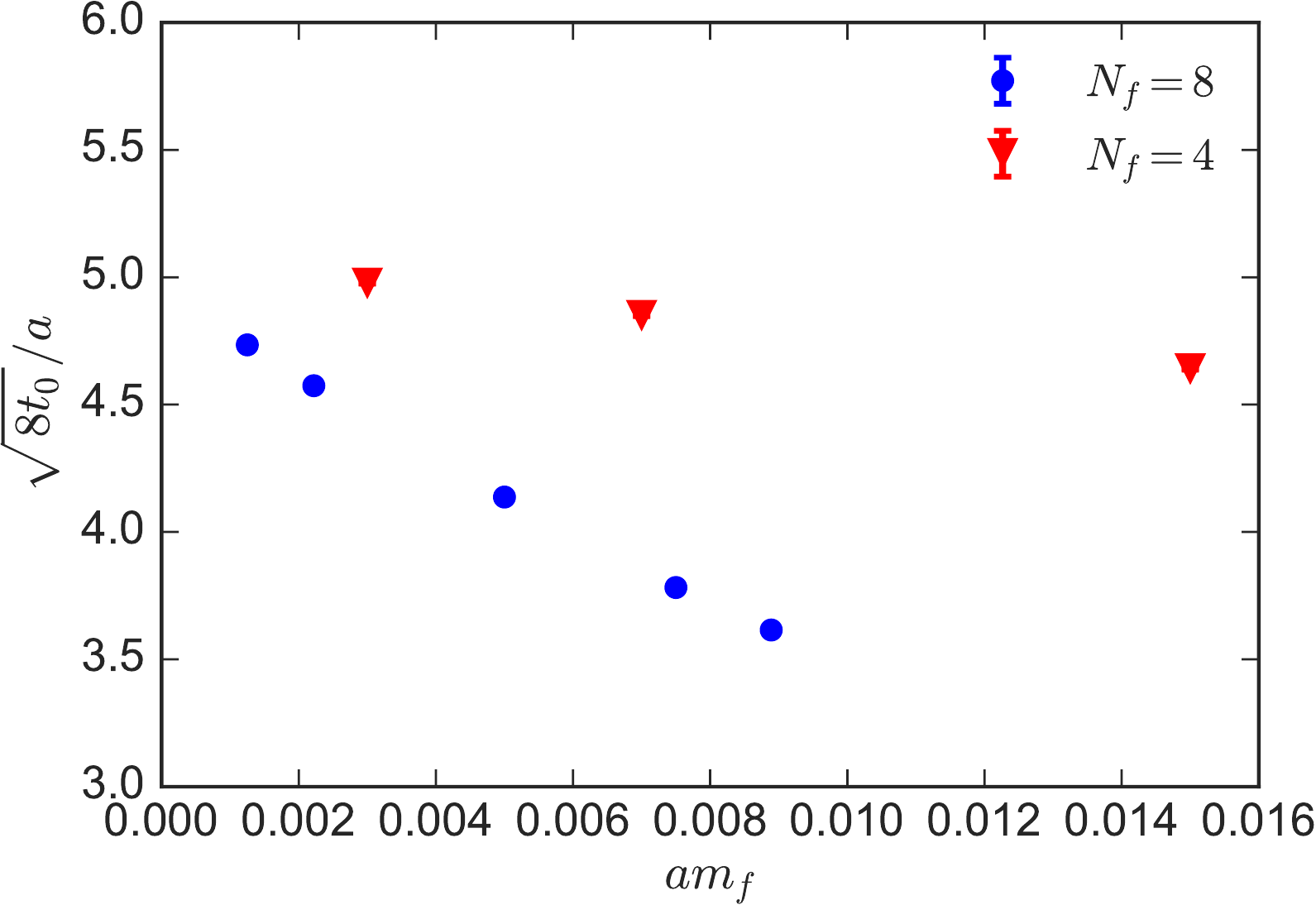}\hfill
  \includegraphics[height=0.21\textheight]{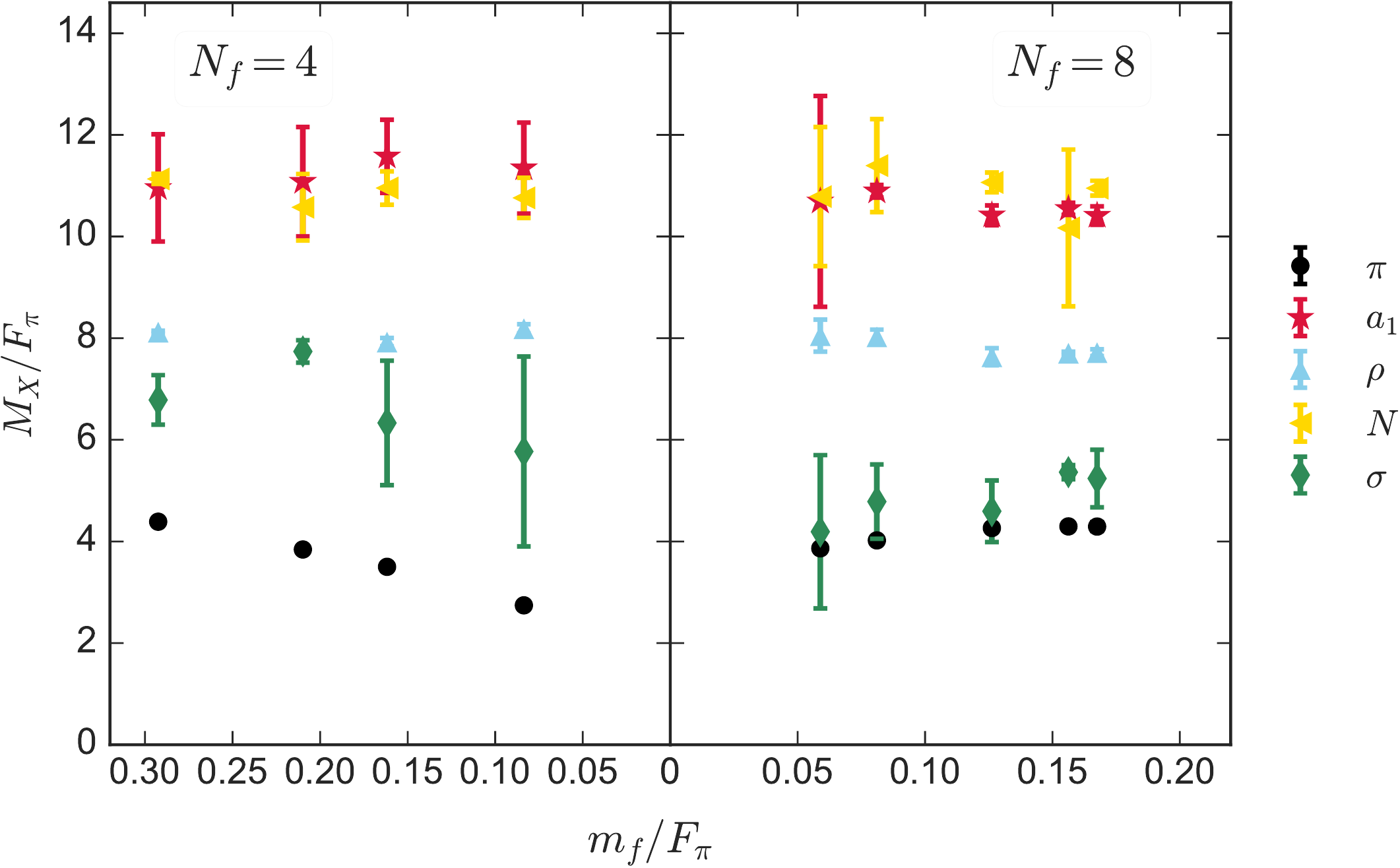}
  \caption{Comparison of SU(3) gauge theories with four and eight flavors by the LSD collaboration from Ref.~\cite{Appelquist:2018yqe}. On the left, the mass dependence of the Wilson flow scale is shown which is larger for $N_f=8$ compared to $N_f=4$. On the right, the spectrum in units of $F_\pi$ is compared showing rather similar values with the exception of the iso-singlet scalar ($\sigma$) which is closer to the rho for $N_f=4$ but degenerate with the pion in $N_f=8$.}
  \label{fig.SU3_8f_4f}
\end{figure}

The differences/similarities between QCD-like simulations with four flavors and near-confor\-mal eight flavors are shown in Fig.~\ref{fig.SU3_8f_4f} where the LSD collaboration compares in the plot on the left the mass dependence of the Wilson flow scale $\sqrt{8t_0}$ between both systems and in the plot on the right the spectrum in units of $F_\pi$ of the low-lying states. For eight flavors $\sqrt{8t_0}$ exhibits a much larger mass dependence than for four flavors which can be interpreted as an effect due to the proximity of an IRFP. The spectrum on the other hand shows surprisingly little differences. All states formed by quark-line connected diagrams are roughly the same in units of $F_\pi$, maybe the ``chiral curvature'' in the pion is less pronounced in eight compared to four flavors. Significantly different is the iso-singlet scalar $0^{++}$ ($\sigma$). For $N_f=4$, its mass is close to the rho and the very large uncertainty at the lightest mass may be an indication that a decay channel opens and the state becomes unstable. In contrast to that, for $N_f=8$, the $\sigma$ is always close to or degenerate with the pion. Since chiral perturbation theory ($\chi$PT) assumes there is only one lightest state (the pion) and all other states are heavier, $\chi$PT cannot provide a valid description of the $N_f=8$ data. This triggered explorations of alternative effective field theories which will be discussed in the following.

\subsection{Recent developments in effective field theories to include a light scalar}
\label{Sec.EFT}

The presence of a light $0^{++}$ scalar in near-conformal gauge theories triggered attention also outside the lattice field theory community. Observing a scalar state as light as the pseudoscalar, rules out to use standard chiral perturbation theory ($\chi$PT) to describe the low energy dynamics. Hence alternative frameworks are considered to obtain an effective field theory (EFT) description. While it is beyond the scope of this review to survey all theories considered, we aim to highlight in the following a selection of recent work. 

\subsubsection{A bound state model for a light scalar }
Relating the numerically observed light $0^{++}$ to near-conformal dynamics, Holdom and Koniuk infer that the light scalar enters the low energy description as a fluctuation around the vacuum expectation value of a scalar doublet \cite{Holdom:2017wpj}. In such a theory, the dynamical fermion mass acts as an order parameter and hence both, pseudoscalar and scalar, are not only degenerate in mass but also have similar couplings to the heavy fermions i.e.~their form factors are similar, too. Using a model inspired from the QED Hamiltonian, they deduce that near-conformal dynamics extending over a relatively wide range of scales leads to a light scalar, well separated from heavier states. Further their model predicts, that in the UV pseudoscalar and scalar form factors are very similar and insensitive to the dynamical fermion mass. This gives rise to a parity-doubled limit of the spin-0 sector with characteristics of a light scalar which allow to distinguish it from a light dilaton. 

\subsubsection{Dilaton effective field theory}
Interpreting the light iso-singlet scalar as a dilation, i.e.~a particle arising from spontaneous breaking of the the conformal symmetry, Appelquist, Ingoldby, and Piai derive an effective field theory where they treat the dilation together with the pions which arise from the spontaneous breaking of the chiral symmetry \cite{Appelquist:2017wcg}. In their model, they add a general form for the dilaton potential and let numerical lattice data determine the parameters of the potential. Using their EFT as fit-ansatz, they conclude that the spectral data for $N_f=8$ with fermions in the fundamental representation as well as $N_f=2$ with sextet fermions can be described by the dilaton EFT \cite{Appelquist:2017vyy}.

\subsubsection{Dilaton-pion low-energy effective theory}

Extending the idea of the dilation EFT, Golterman and Shamir investigate the Veneziano limit i.e.~they consider the limit of $N_f \to \infty$ for $N_f /N_c$ fixed \cite{Golterman:2018mfm,Golterman:2018bpc}. Denoting by $n_f^*$ the onset of the conformal window in the Veneziano limit, they expand around $n_f - n_f^*$  with $n_f = N_f /N_c$. This expansion allows to identify two regions. In the \textbf{small mass region}, the dilaton decouples from the pions and a typical chiral behavior can be observed. In the \textbf{large mass region}, hadron masses and decay constants exhibit hyperscaling proportional to $m_f^{1/(1+\gamma^*)}$, where $m_f$ is the fermion mass and $\gamma^*$ the anomalous dimension.

Testing the implications of the dilaton-pion low-energy EFT, Golterman and Shamir conclude that LSD's $N_f=8$ data are in the large mass region and provide an explanation for some of the observed characteristics. In order to reach the small mass region, $m_f$  would need to be reduced to about $m_f/100$. Investigating that region may reveal that $N_f=8$ is indeed confining. A slightly different conclusion regarding the mass region of LSD's $N_f=8$ data is reached in Ref.~\cite{Fodor:2019vmw} where also dilaton EFT fits to LatHC's sextet model data are presented.

\subsubsection{Linear sigma model for multiflavor gauge theories}

A different ansatz inspired the by linear sigma model \cite{GellMann:1960np} is proposed by Meurice. In addition to the pion and $\sigma$ ($0^{++}$) also the $a_0$ and the $\eta^\prime$ are considered to derive an effective theory for gauge theories with many flavors \cite{Meurice:2017zng}. This ansatz investigates the role of the explicit breaking of the axial $U_A(1)$ symmetry on the spectrum and how it depends on the number of flavors $N_f$ near the onset of conformal window. At tree-level, he derives relations for dimensionless ratios
\begin{align}
R_\sigma &= (M_\sigma^2 -M_\pi^2)/M_{\eta^\prime}^2 + (1-2/N_f)(1 - M_\pi^2/M_{\eta^\prime}^2) \nonumber \\
R_{a_0} &= (M_{a_0}^2 -M_\pi^2)/M_{\eta^\prime}^2 - (2/N_f)(1 - M_\pi^2/M_{\eta^\prime}^2),
\end{align}
which he tests using  LatKMI data for $N_f=8$ and 12. The data show an almost flat behavior and no $N_f$ dependence. This may allow to derive e.g.~a bound on $N_{fc}$, the critical number of flavors denoting the onset of the conformal window.

\subsubsection{Mass splittings in a linear sigma model for multiflavor gauge theories}
Subsequently Meurice together with DeFloor and Gustafson extended his linear sigma model to the class of mass-split systems \cite{DeFloor:2018xrp,deFloor:2018zdj} discussed in Sec.~\ref{Sec.mass-split}. Mass-split models have two type of flavors, light flavors with mass $m_1$ and heavy flavors of mass $m_2$. A particle spectrum made-up from light and heavy flavors exhibits light-light, heavy-light, and heavy-heavy mesons. Considering the case where both masses are similar, differing only by a small amount $\delta_m$, i.e.~$m_2 = m_1 + \delta_m$, they derive a peculiar ordering of states.  If the pseudoscalars exhibit the expected ``normal'' order $M^2_{\pi ll} < M^2_{\pi hl} < M^2_{\pi hh}$, then inverse ordering is predicted for the scalars $M^2_{a_0 ll} > M^2_ {a_0 hl} > M^2_{a_0 hh}$.

Available data from simulations with four light and eight heavy flavors \cite{Brower:2014dfa,Brower:2015owo,Hasenfratz:2016gut,Hasenfratz:2017lne,Witzel:2018gxm} are not performed with similar flavor masses. These data do not exhibit the predicted inverse ordering.

\subsubsection{Linear sigma EFT for nearly conformal gauge theories}
Motivated by their numerical results of simulations with $N_f=8$ dynamical flavors, the LSD collaboration seeks a description to accommodate a light iso-singlet scalar as well as a pseudoscalar decay constant varying significantly with the bare fermion mass \cite{Appelquist:2018yqe}. Point of origin for their derivation is the linear sigma model because the scalar potential in the linear sigma model breaks chiral symmetry spontaneously and leads to a tree-level quark mass dependence \cite{Appelquist:2018tyt}. In contrast to Ref.~\cite{Meurice:2017zng} discussed above, the $\eta^\prime$ is assumed to be heavy due to mixing with topological fluctuations and integrated out by hand. Subsequently they deduce tree-level expressions for the masses of the pion, $\sigma$, and $a_0$ away from the chiral limit as a function of the scalar potential. Introducing a spurion field to represent the explicit breaking of chiral symmetry coming from the quark mass, they organize the contributing operators and obtain a relation between $M_\pi$ and $M_\sigma$. For small chiral symmetry breaking this relation predicts $M_\sigma^2 \ge 3 M_\pi^2$, which is incompatible with present lattice data. However, this relation is relaxed for sufficiently large chiral symmetry breaking. Hence the linear sigma EFT may provide a viable description e.g.~of the $N_f=8$ data.

\section{Higgs as a pseudo Nambu-Goldstone boson}
\label{Sec.Higgs_pNGB}
We discuss three examples for composite Higgs models assuming the Higgs to be a pNGB. First we introduce a project targeting models based on the SU(4)/Sp(4) coset. Next we summarize and present numerical results for ``Ferretti's model'' constructed with fermions in two representations and end this section by discussing mass-split models.
\subsection{Composite Higgs model based on the SU(4)/Sp(4) coset}

A pNGB composite Higgs model based on the SU(4)/Sp(4) coset is favored by some phenomenological considerations (see e.g.~\cite{Katz:2005au,Gripaios:2009pe,Barnard:2013zea,Cacciapaglia:2015eqa}). Since this coset emerges naturally in gauge theories with pseudo-real representations, one possibility is to investigate Sp(2N) gauge theories with two massless, fundamental Dirac flavors. To address the lack of knowledge on Sp(2N) gauge theories, Bennett, Hong, Lee, Lin, Lucini, Piai, and Vadacchino started a larger program to investigate such gauge theories for $N_f=2$ fundamental flavors and $N>1$.

\begin{figure}[tb]
  \centering
  \includegraphics[height=0.22\textheight]{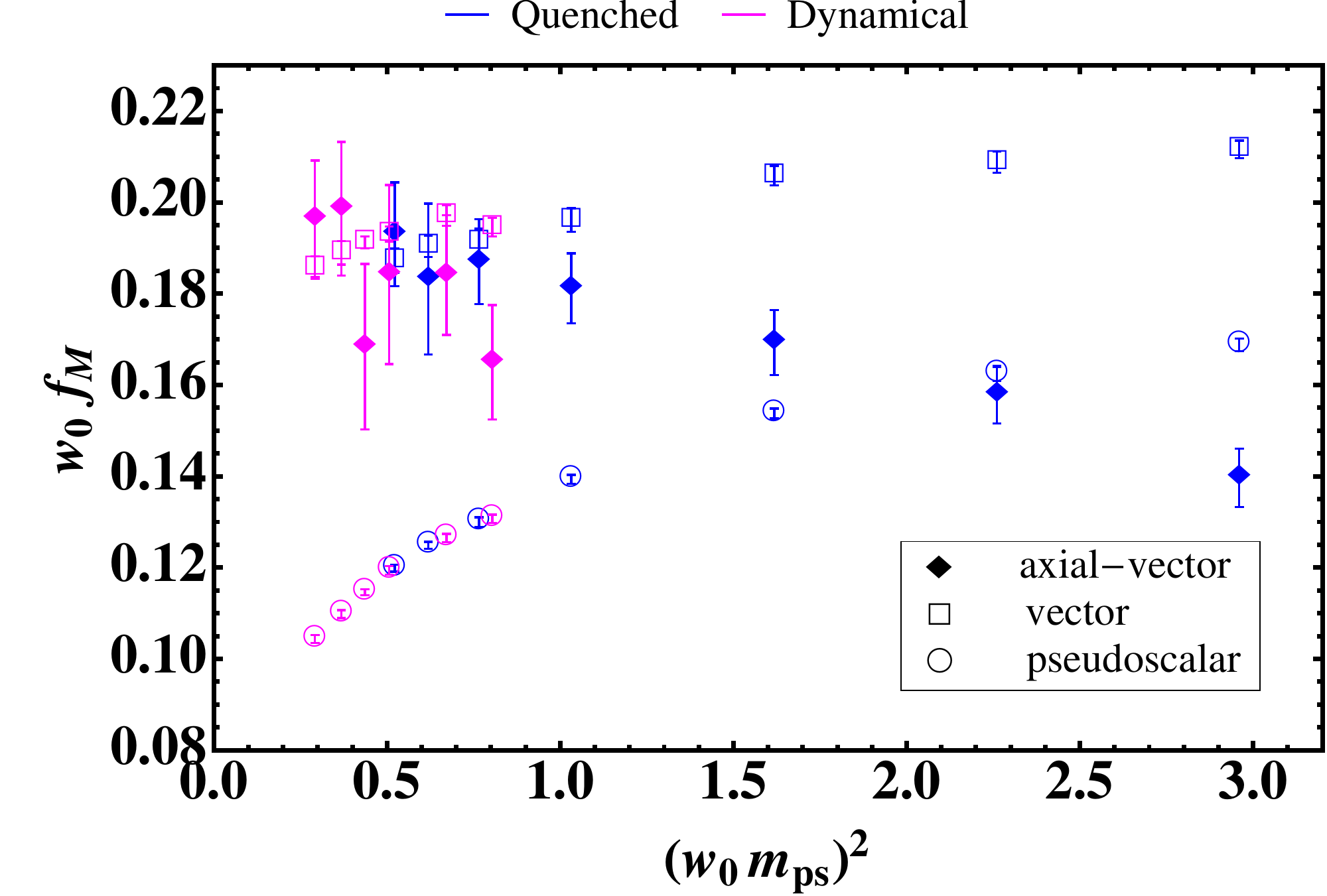}\hfill
  \includegraphics[height=0.22\textheight]{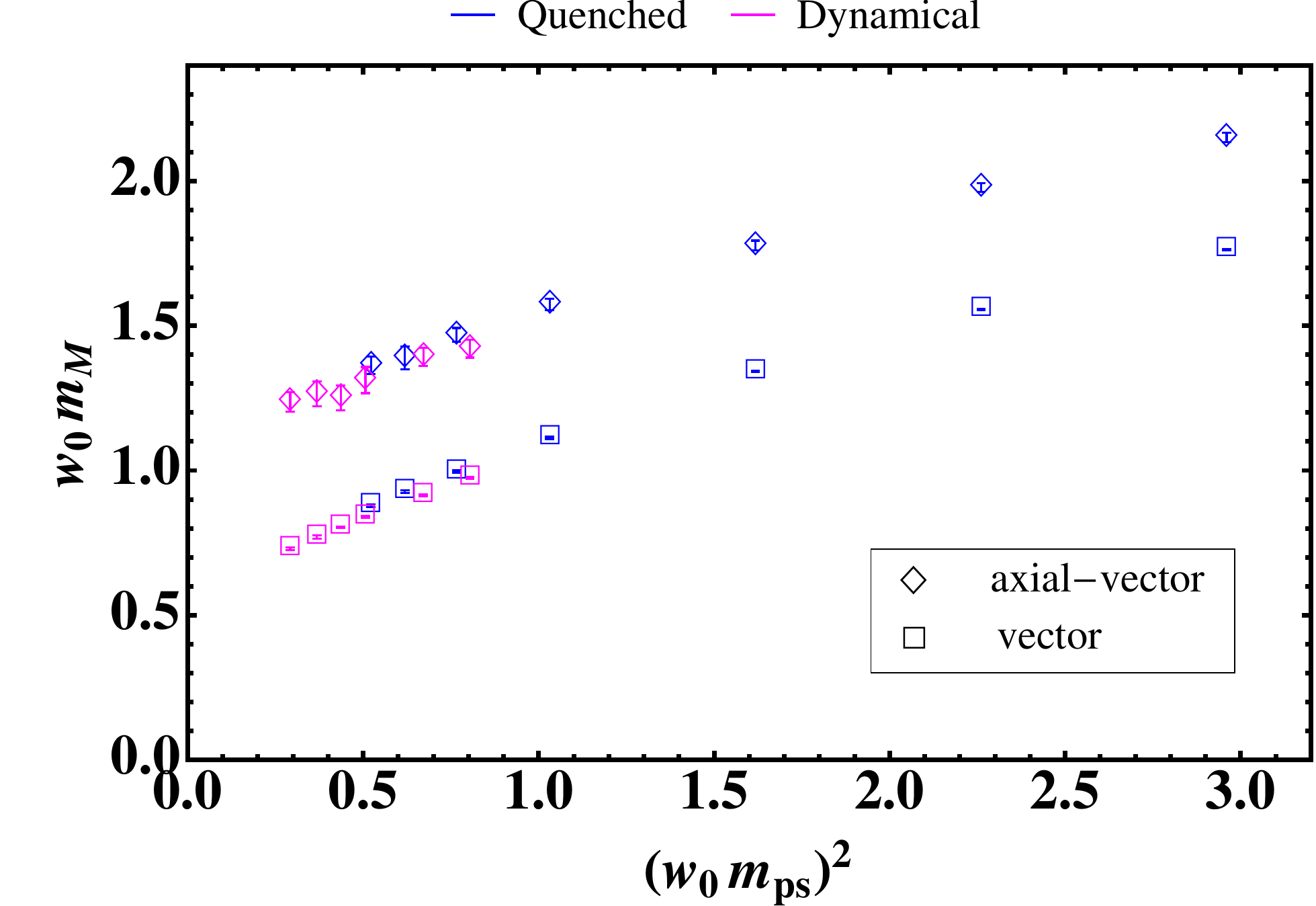}
  \caption{Comparison of results obtained from quenched and dynamical simulations for an Sp(4) gauge theory with two fundamental Dirac flavors. Shown as a function of the squared pseudoscalar mass are the pseudoscalar, vector, and axial-vector decay constants on the left; vector and axial-vector mass on the right. All quantities are presented in units of the $w_0$ Wilson flow scale. Plots courtesy by Lee \cite{Lee:2018prv}.}
  \label{fig.SU4_SP4}
\end{figure}

This work extends published results predominantly obtained in the quenched limit using an Sp(4) gauge theory \cite{Bennett:2017kga} by performing dynamical simulations with Wilson fermions. Figure \ref{fig.SU4_SP4} presents first dynamical results \cite{Lee:2018prv} for masses and decay constant in comparison to quenched data. As the plot shows, qualitative agreement between quenched and dynamical results is observed.

\subsection{Ferretti’s Model}
Ferretti proposed a model \cite{Ferretti:2014qta} designed to exhibit a Higgs boson as pNGB and explains the large top quark mass via partial compositeness i.e.~the top quark mixes with a composite state of the new strong sector which has the same quantum numbers. The model is based on an SU(4) gauge theory and exhibits fermions in two representations:
\begin{itemize}
\item $N_6^W = 5$ massless Weyl flavors of sextet fermions in the two-index antisymmetric representations denoted by $Q$ and carrying electro-weak charges
\item  $N_4 = 3$ fundamental Dirac flavors with color charge and denoted by $q$.
\end{itemize}
Due to fermions with two representations, this systems exhibits a rich spectrum. Mesons can be created as a pair of sextet flavors, $QQ$, $Q \bar Q$, $\bar Q \bar Q$ or as a pair of fundamental flavors $q\bar q$. In both cases pNGBs and vector states arise. Baryonic matter are either sextet bosons ($QQQQQQ$), fundamental bosons ($qqqq$), or so called chimera fermions $Qqq$. Of phenomenological interest is the ``Ferretti limit'' for which the mass of the sextet flavors goes to zero ($m_6 \to 0$). In that limit the Higgs boson is a massless sextet NGB and its potential arises from SM interactions. SM fermions acquire its mass from quartic mixing $u\bar uH \to u\bar uQQ$. The large top quark mass is a consequence of the linear mixing of the top quark with the chimera. Furthermore, this system exhibits a non-anomalous superposition of $U_{A(4)}(1)$ and $U_{A(6)}(1)$. This leads to an axial singlet pNGB, referred to as $\zeta$ meson.\\

Numerical investigations of Ferretti's model have been carried out by the TACoS collaboration \cite{Ayyar:2017qdf,Ayyar:2018zuk,Ayyar:2018ppa}. To simplify the simulations, they adapted Ferretti's model to a more conventional lattice field theory set-up. Their simulations are based on an SU(4) gauge theory with $N_6 = 2$ Dirac flavors (corresponding to $N^W_6$ = 4 Weyl flavors) in the sextet representation and $N_4 = 2$ fundamental Dirac flavors. In the following we briefly summarize some of their results.

\begin{figure}[tb]
  \centering
   \includegraphics[height=0.25\textheight]{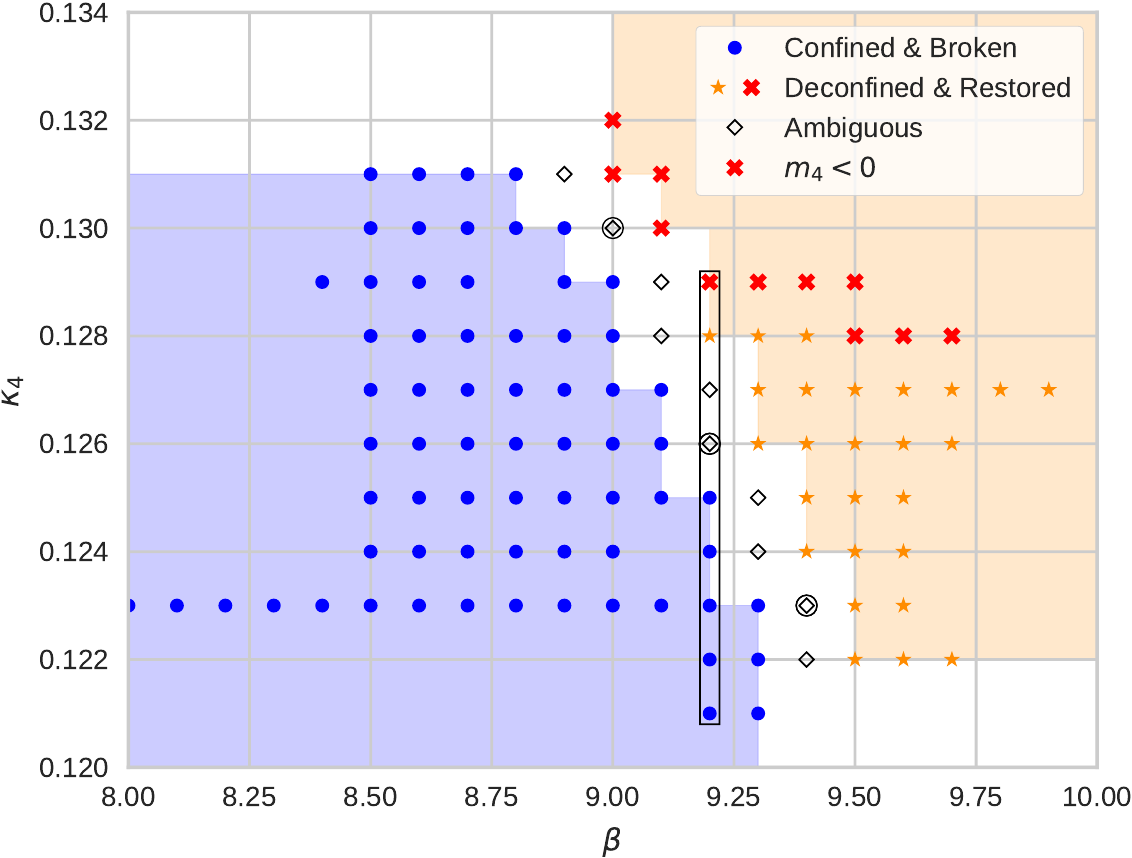} \hfill
   \includegraphics[height=0.25\textheight]{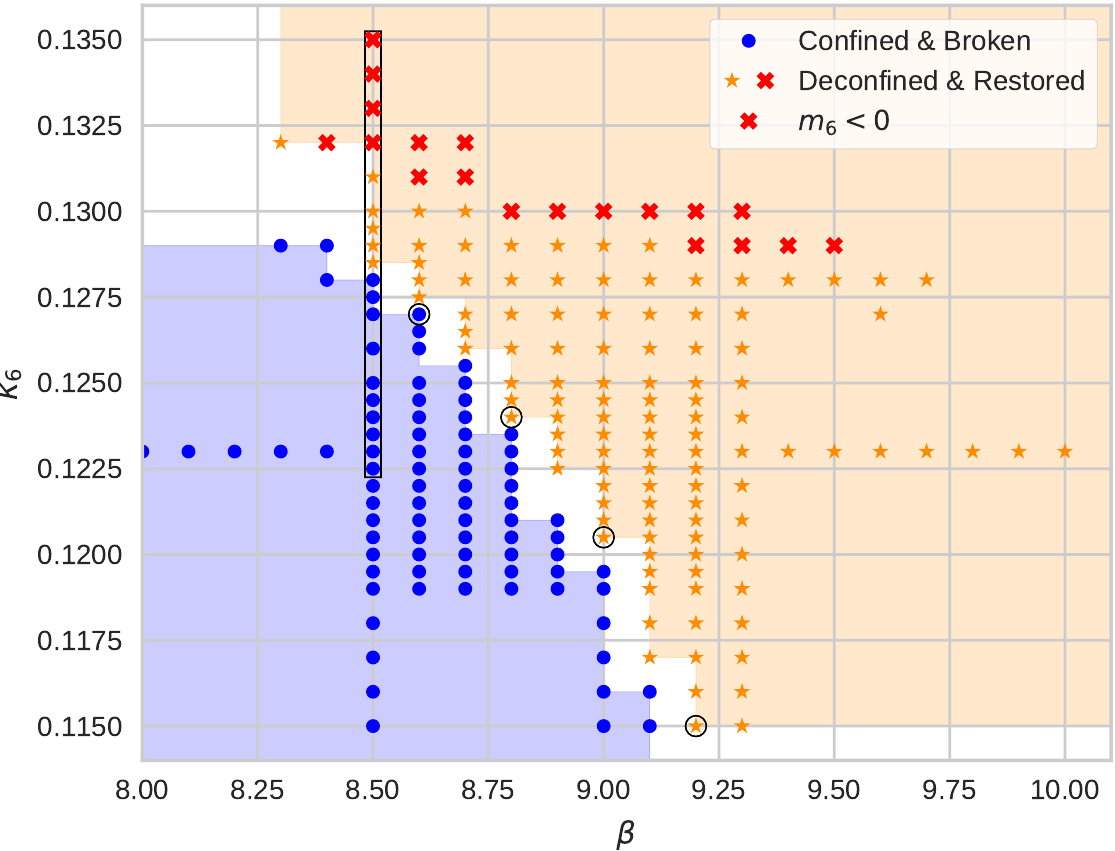}
   \caption{Finite temperature phase diagram of the adapted Ferretti model from Ref.~\cite{Ayyar:2018ppa}. Numerical investigations identify two phases. As theoretically predicted the transition appears to be first order. }
   \label{fig.Ferretti_finiteT}
\end{figure}

\textbf{Finite temperature phase diagram}: Exploring the phase diagram in dependence of the three parameters $\beta$, $\kappa_6$, and $\kappa_4$, only two phases are identified as shown in Fig.~\ref{fig.Ferretti_finiteT}. A low-temperature phase where both fermion species are confined and chirally broken as well as a high-temperature phase with both fermion species deconfined and chirally restored. The single phase transition appears to be first order as theoretically predicted \cite{Ayyar:2018ppa}.

\begin{figure}[tb]
  \includegraphics[height=0.25\textheight]{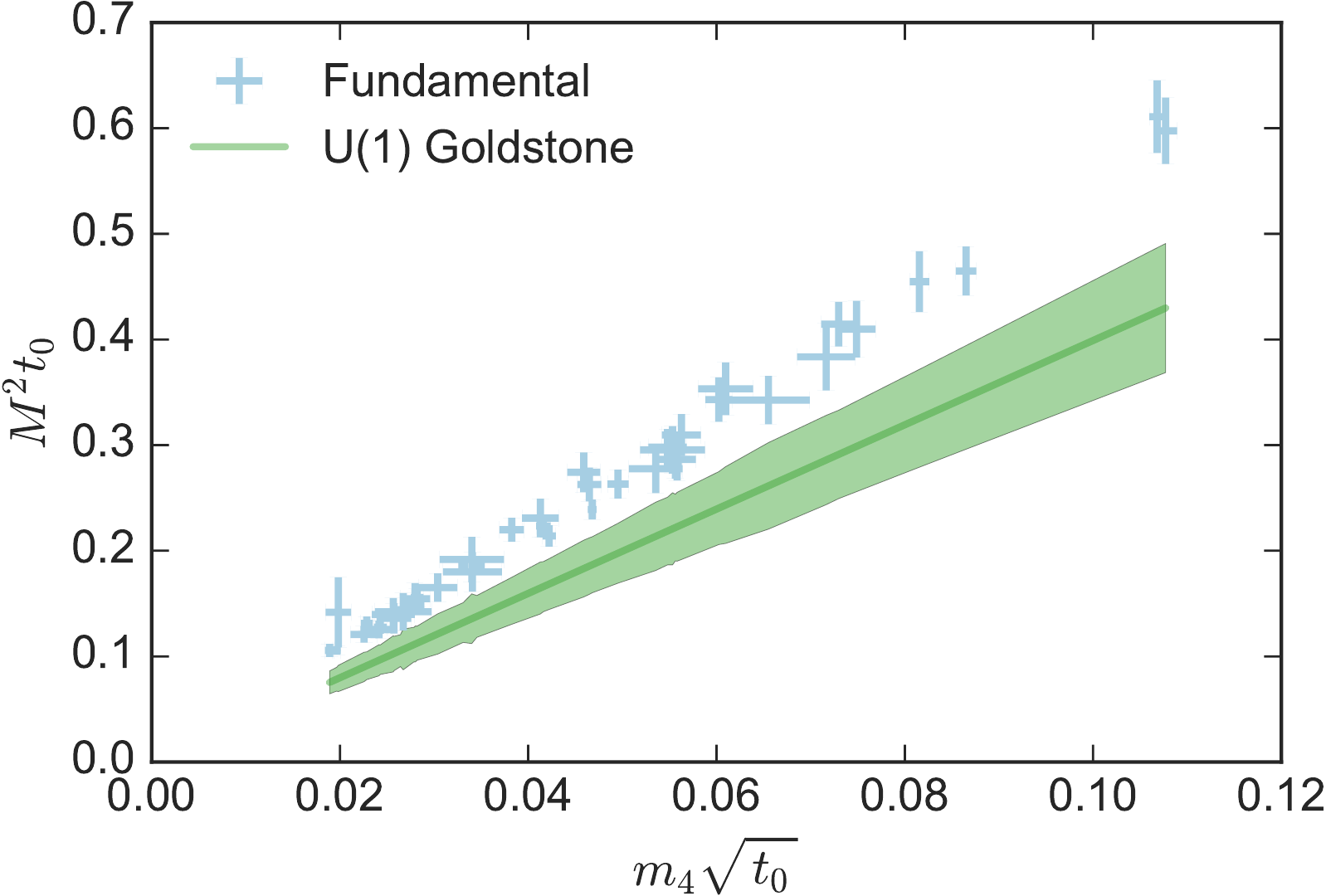}\hfill
  \includegraphics[height=0.25\textheight]{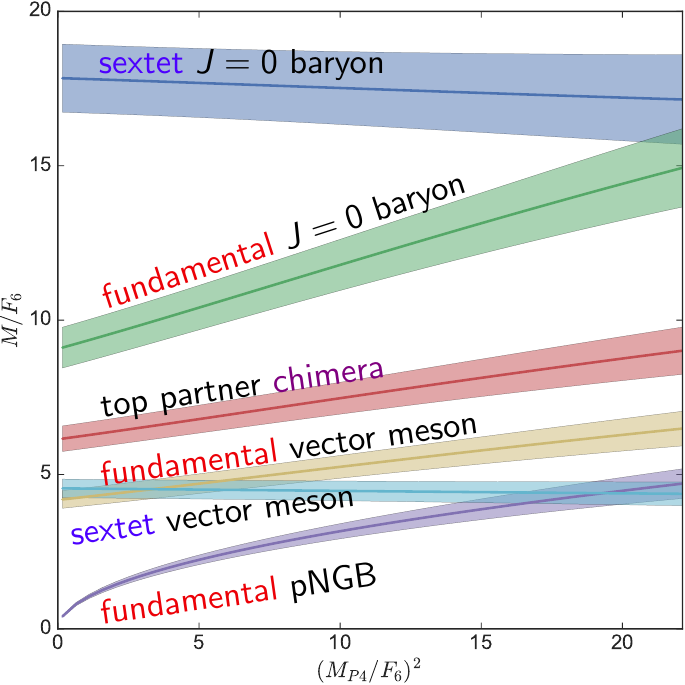}
  \caption{Left: Mass dependence of the $\zeta$-meson reconstructed from a chiral fit as function of $m_4$ and $m_6$ from Ref.~\cite{Ayyar:2017qdf}. Right: overview of the spectrum in the Ferretti limit ($m_6\to 0$) from Ref.~\cite{Ayyar:2018zuk}.}
  \label{fig.Ferretti_spectrum}
\end{figure}

\textbf{Spectrum results:} Using zero-temperature simulations the spectrum of the adapted Ferretti model is explored. The left panel in Fig.~\ref{fig.Ferretti_spectrum} shows the $\zeta$ meson which is reconstructed from a chiral fit performed as function of $m_4$ and $m_6$. Since in the Ferretti limit ($m_6 \to 0$), the sextet pNGBs are exactly massless ($M_{PS6}=0$), it turns out that the $\zeta$ meson is the lightest massive state, lighter than the pNGBs of the four fundamental flavors ($M_\zeta<M_{PS4}$). The right plot in Fig.~\ref{fig.Ferretti_spectrum} presents an overview of the spectrum after taking the $m_6\to 0$ limit. Results are presented in units of the sextet pseudoscalar decay constant, $F_6$, as a function of the four flavor pseudoscalar mass squared. The most recent work \cite{Ayyar:2018glg} addresses mechanisms for partial compositeness in this model.

\subsection{Mass-split models}
\label{Sec.mass-split}
A generic framework to explore composite Higgs scenarios with near-conformal dynamics is given by mass-split models. Deduced from the understanding that promising models are chirally broken in the IR but conformal in the UV \cite{Luty:2004ye,Dietrich:2006cm,Vecchi:2015fma,Ferretti:2013kya}, mass-split models use flavors with different masses to generate the desired dynamics. As an example, we start with an SU(3) gauge theory and add ``heavy'' and ``light'' (massless) fundamental flavors. The number of light flavors $N_\ell$ is chosen such that a system with only $N_f=N_\ell$ flavors is chirally broken in the IR. Next we choose $N_h$ heavy flavors to push the system near an IRFP of a conformal theory i.e.~we choose $N_h$ such that a theory with $N_\ell+N_h$ degenerate flavors is conformal.
In the case of SU(3) with $N_\ell=4$ light flavors, the model can be embedded in a fundamental composite two Higgs doublet model (2HDM) \cite{Ma:2015gra,BuarqueFranzosi:2018eaj}. Moreover, the heavy flavors could be invisible to the SM. Hence experiments may only observe light-light and heavy-light states.

Mass-split models have unique properties because they combine features of chirally broken (QCD-like) and conformal systems. Let us first recall how to take the appropriate limits in the limiting cases of a chirally broken or a conformal system, respectively. In QCD, e.g., the continuum limit is taken by sending both, the gauge coupling and the fermion mass, to zero, $g^2,\,m_f\to 0$. In the case of a (mass-deformed) conformal theory with degenerate $N_f$ flavors the system exhibits an IRFP. Hence the continuum limit is taken by sending the fermion mass $m_f \to 0$ whereas the gauge coupling is an irrelevant parameter. As consequence, all ratios of hadron masses scale with the anomalous dimension, a property referred to as hyperscaling. 

Since mass-split models live in the basin of attraction of the IRFP corresponding to $N_f$ degenerate flavors, they inherit hyperscaling of ratios of hadron masses but due to splitting the masses, these models are by construction chirally broken. The continuum limit is obtained for $m_h \to 0$ keeping the ratio of flavor masses $m_\ell/m_h$ fixed. The chiral limit is however approached for $m_\ell \to 0$ i.e.~also the ratio $m_\ell/m_h \to 0$. Like for a conformal system, the gauge coupling in mass-split systems is an irrelevant parameter. Hence after taking the chiral and continuum limit, there is no free parameter but a highly constrained spectrum with light-light, heavy-light, and heavy-heavy bound states \cite{Hasenfratz:2016gut}.
In the following we demonstrate the unique properties of mass-split systems by presenting results for an SU(3) gauge theory with four light and eight heavy flavors obtained by Hasenfratz, Rebbi, and Witzel using simulations with nHYP-smeared staggered fermions.

\begin{figure}[tb]
  \centering
  \includegraphics[height=0.30\textheight]{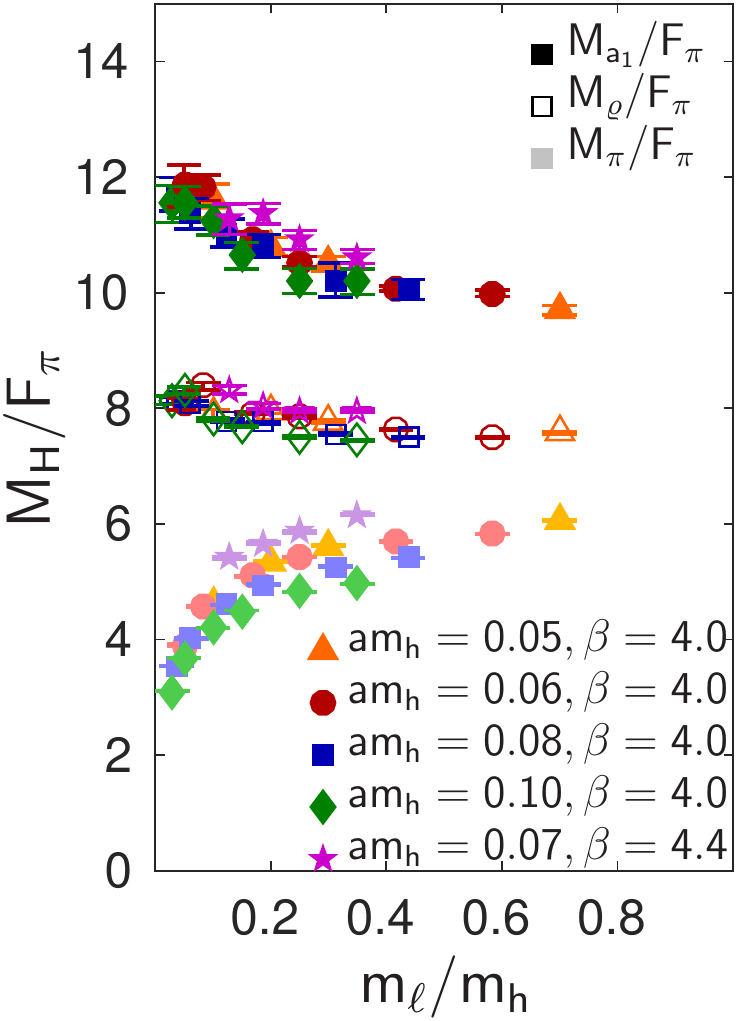}
  \includegraphics[height=0.30\textheight]{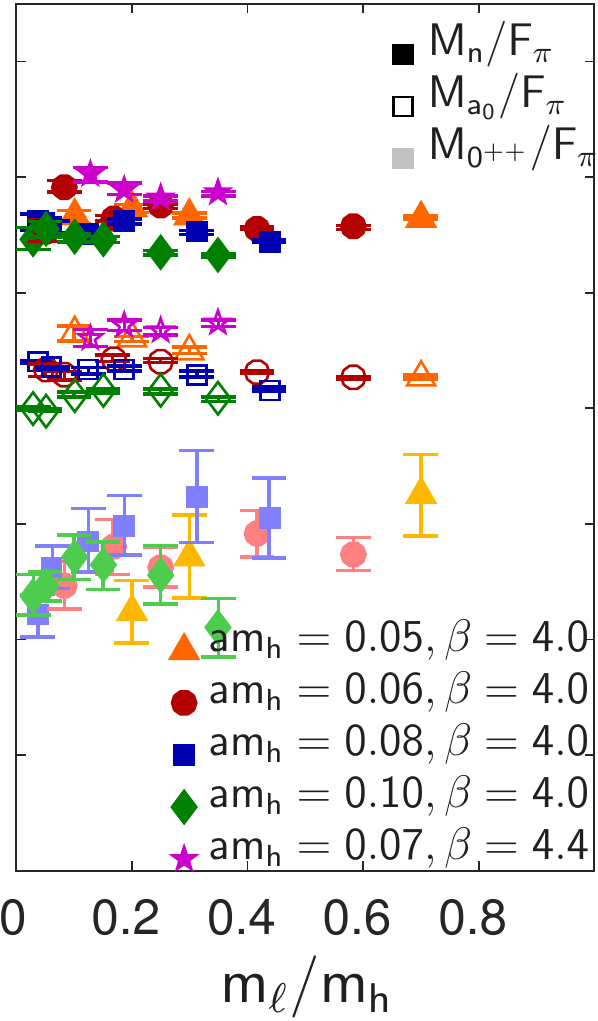}
  \caption{Light-light spectrum in units of $F_\pi$ from Ref.~\cite{Brower:2015owo}. Up to small scaling violations, unique curves independent of the heavy flavor mass $m_h$ and the bare gauge coupling $\beta$ are observed for all states. The iso-singlet scalar ($0^{++}$) is degenerate with the pseudoscalar ($\pi$) and much lighter than the vector ($\varrho$).}
  \label{fig.fpe_hyperscaling}
\end{figure}

\textbf{Light-light spectrum:} Figure \ref{fig.fpe_hyperscaling} shows the pseudoscalar ($\pi$), vector ($\rho$), axial ($a_1$), iso-singlet scalar ($0^{++}$), multiplet scalar ($a_0$), and the nucleon (n) formed only from light flavors in units of the pseudoscalar decay constant ($F_\pi$) plotted vs.~the ratio of flavor masses $m_\ell/m_h$. For all states and up to small scaling violation effects, the ratios of hadron masses $M_H$ over $F_\pi$ trace out unique curves depending only on $m_\ell/m_h$. The heavy flavor mass $m_h$ sets the scale and the gauge coupling is an irrelevant parameter. Further, the  iso-singlet scalar is found to be degenerate with the pion and much lighter than the rho.

\begin{figure}[tb]
  \centering
  \includegraphics[height=0.23\textheight]{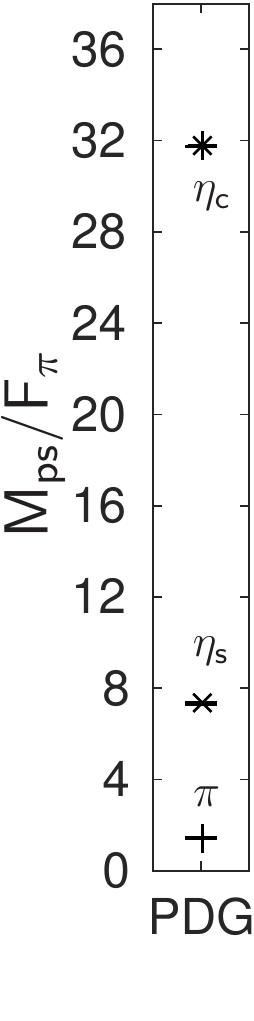}
  \includegraphics[height=0.23\textheight]{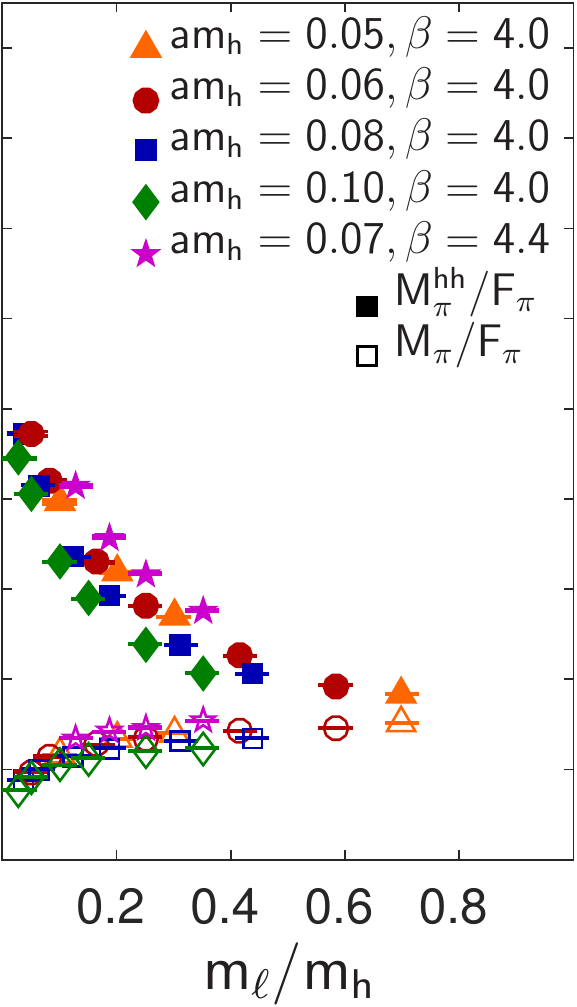}
  \includegraphics[height=0.23\textheight]{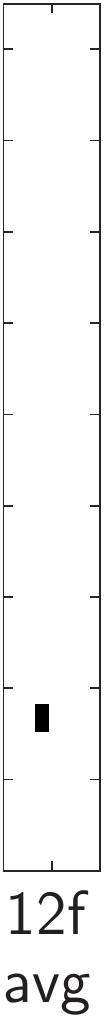}
  \includegraphics[height=0.23\textheight]{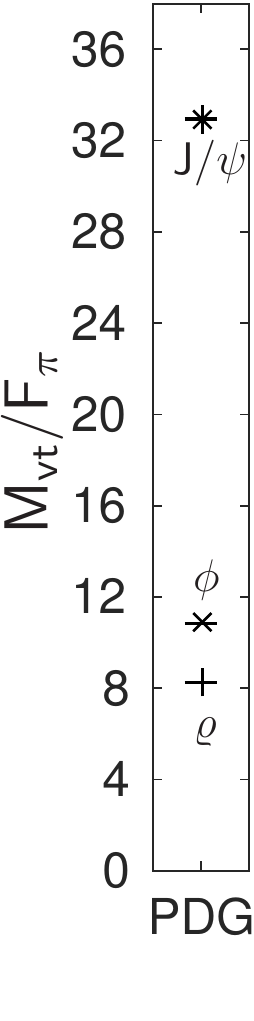}
  \includegraphics[height=0.23\textheight]{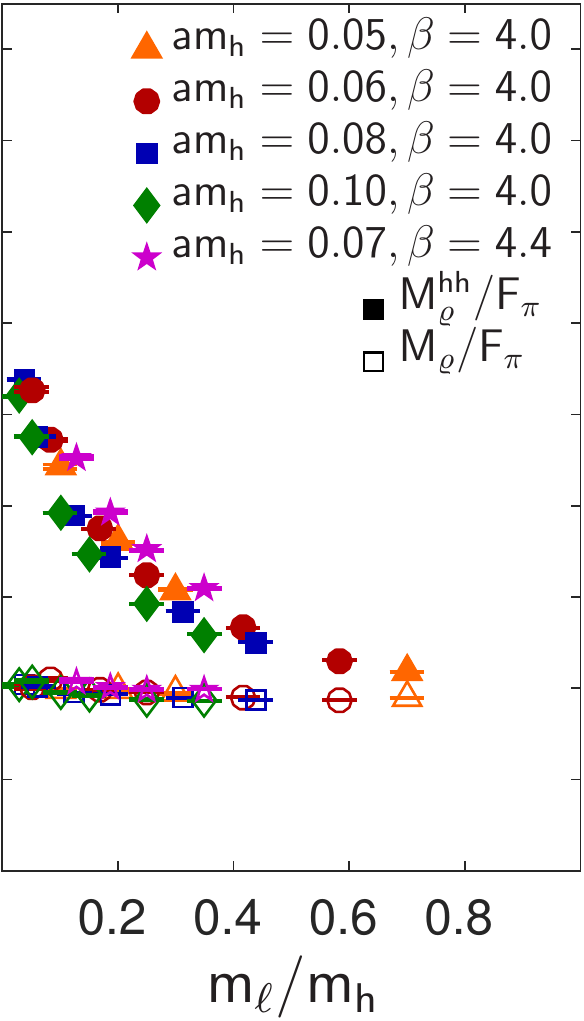}
  \includegraphics[height=0.23\textheight]{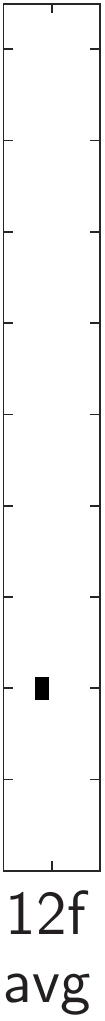}
  \includegraphics[height=0.23\textheight]{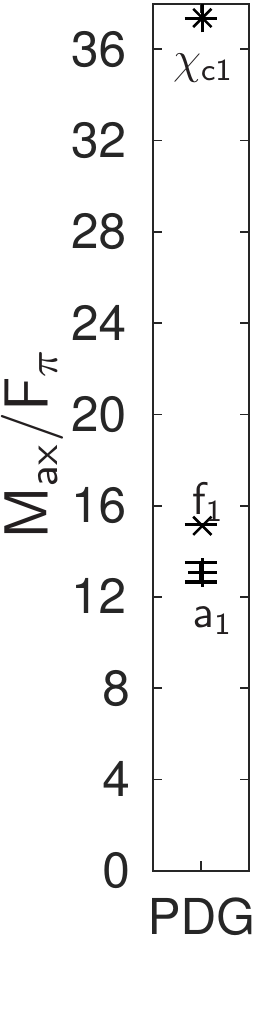}
  \includegraphics[height=0.23\textheight]{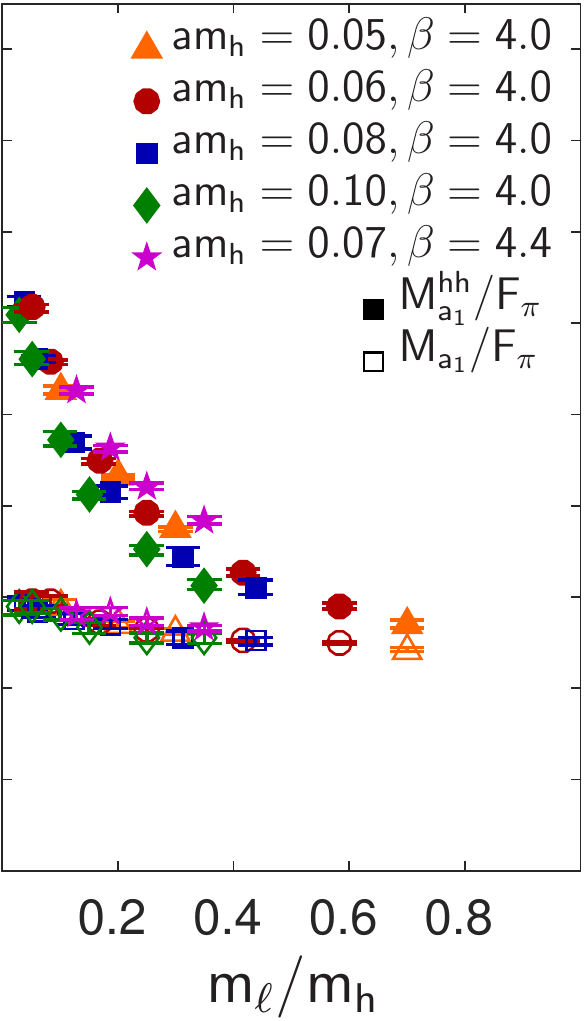}
  \includegraphics[height=0.23\textheight]{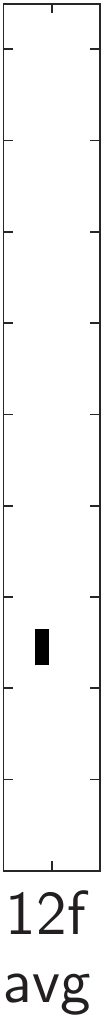}
  \caption{Light-light and heavy-heavy spectrum for pseudoscalar, vector, and axial states from Ref.~\cite{Hasenfratz:2016gut}. Also the heavy quarkonia are independent of $m_h$ and $\beta$ which results in qualitative differences to QCD-like systems.}
  \label{fig.fpe_hh_ll}
\end{figure}

\textbf{Heavy-heavy spectrum:} The same features are present in the heavy-heavy spectrum as can be see in Fig.~\ref{fig.fpe_hh_ll} where pseudoscalar, vector, and axial states of either only light or only heavy flavors are shown. While in the limit of $m_\ell/m_h\to 1$, the limiting case of degenerate 12 flavors is approached. The chiral limit ($m_\ell/m_h \to 0$) differs significantly from QCD. In QCD, the masses of quarkonia are proportional to the constituent quark mass whereas the 4+8 model exhibits only one quarkonia mass independent of the mass of the constituent heavy flavor.

\begin{figure}[tb]
  \centering
  \includegraphics[height=0.17\textheight]{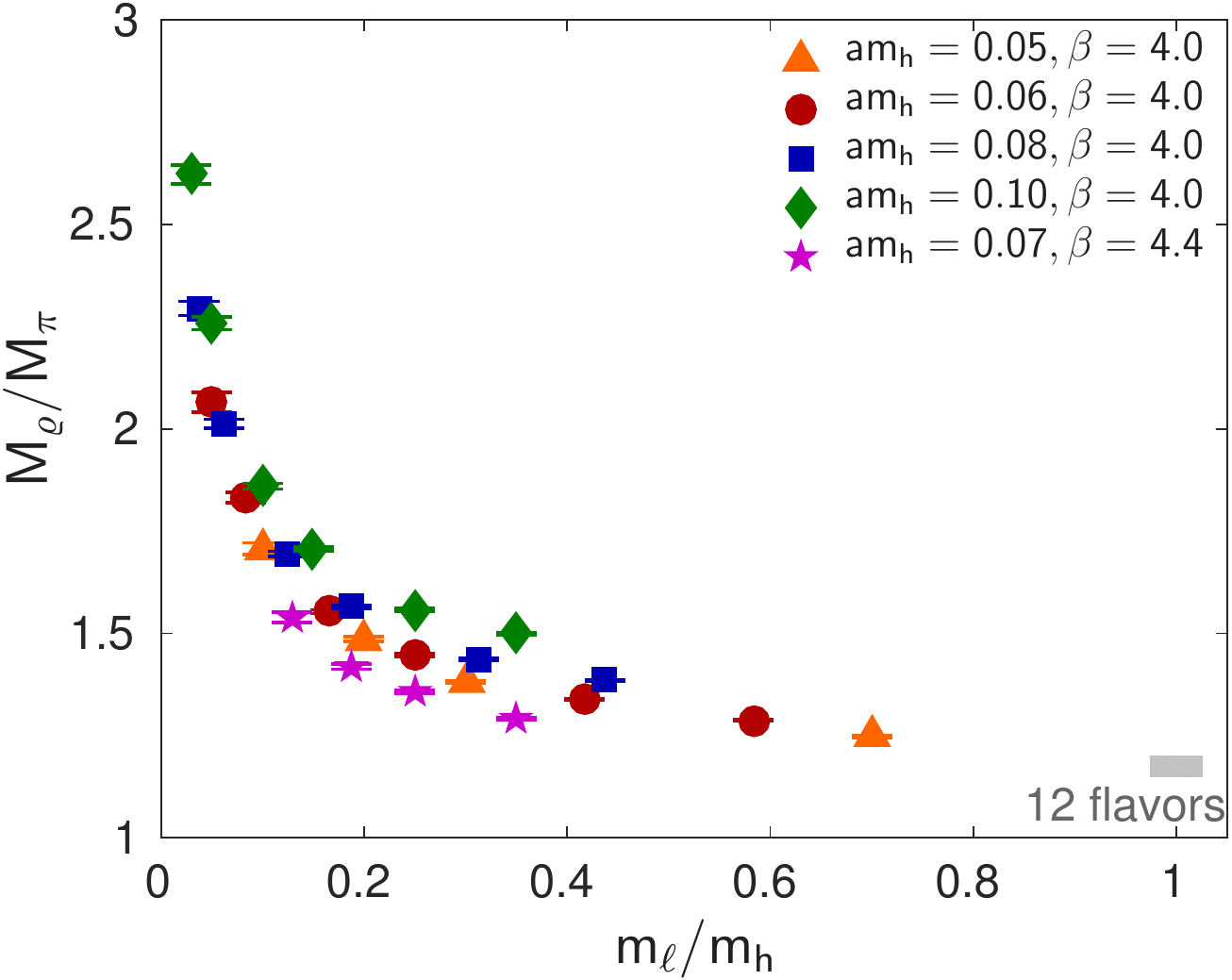}
  \includegraphics[height=0.17\textheight]{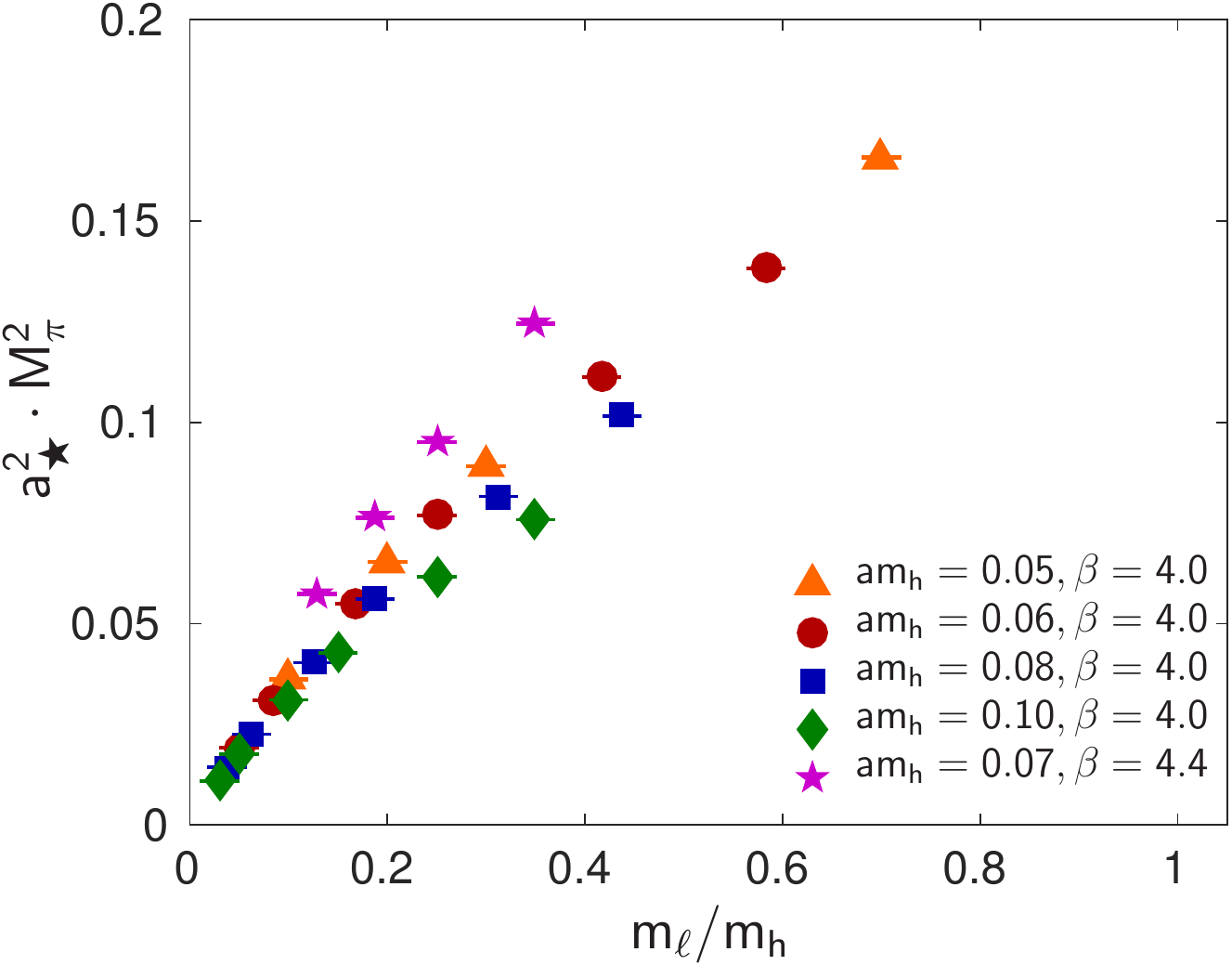}  
  \includegraphics[height=0.17\textheight]{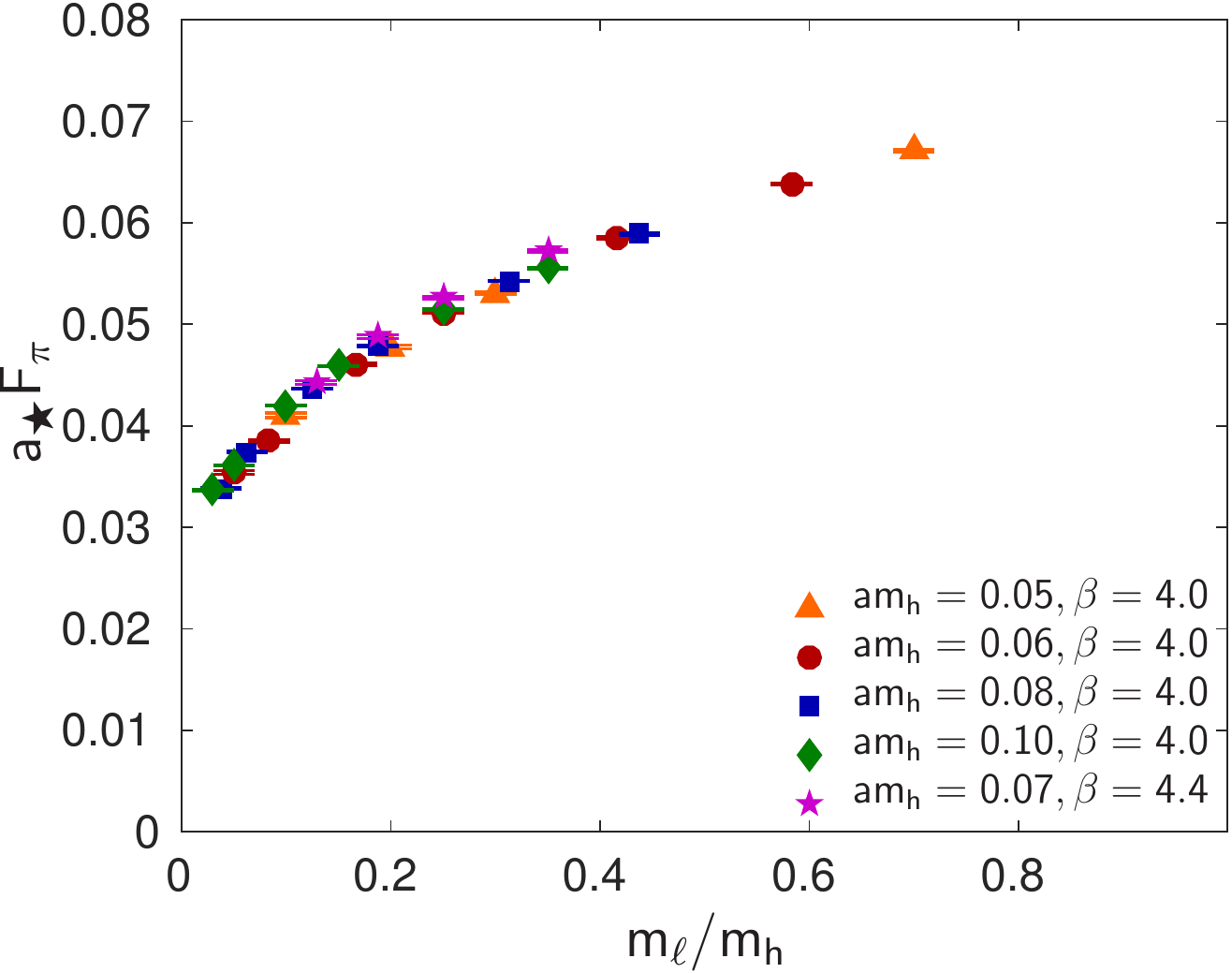}
  \caption{Demonstration of the chirally broken nature of the mass-split model from Refs.~\cite{Hasenfratz:2016gut,Hasenfratz:2017lne,Witzel:2018gxm}. Left: the ratio of $M_\varrho/M_\pi$ diverges in the chiral limit. Center: the squared pion mass scales linearly in $m_\ell/m_h$. Right: the pseudoscalar decay constant $F_\pi$ has a finite value for $m_\ell/m_h\to 0$. }
  \label{fig.fpe_chibroken}
\end{figure}

\textbf{Chirally broken:} To demonstrate that this model is indeed chirally broken, we finally show the diverging ratio of $M_\rho/M_\pi$, the linear dependence of $a_\bigstar M_\pi^2$ on $m_\ell/m_h$, as well as that the decay constant $a_\bigstar F_\pi$ is approaching a finite value for $m_\ell/m_h\to 0$. The notation $a_\bigstar$ indicates that results are converted to the same lattice units using ratios of the Wilson flow scale $\sqrt{8t_0}$. 

\subsection{Future perspectives}
A new large scale investigation by the LSD collaboration is in progress to further explore mass-split models. Using four light and six heavy flavors they observe indications of hyperscaling \cite{Witzel:2018gxm} but more statistics and insight on possible systematic effects is required before reaching conclusions. Hyperscaling in the 4+6 system would indicate that a system with degenerate ten fundamental flavors is indeed (near-)conformal as suggested by Chiu's step-scaling analysis \cite{Chiu:2016uui,Chiu:2017kza,Chiu:2018edw}. Phenomenologically this would be interesting because $N_f = 10$ is then also expected to have a larger anomalous dimension than the 12-flavor system. The numerical investigations of the 4+6 system are carried out using stout-smeared M\"obius domain wall fermions simplifying the calculation of phenomenologically interesting processes and quantities like the generation of mass for SM fermions (partial compositeness, four-fermion interaction), baryon anomalous dimension (e.g.~via new gradient flow method \cite{Carosso:2018bmz}), the $S$-parameter \cite{Appelquist:2010xv}, or the Higgs-potential. Another interesting avenue would be to combine the two representation model with the idea of mass-split systems and push the two representation model closer to the conformal window by adding more flavors.

\section{Further developments}
\label{Sec.further}
\subsection{Dynamical generation of elementary particle masses}

Following a proposal by Frezzotti and Rossi \cite{Frezzotti:2014wja}, Capitani, Divitiis, Dimopoulos, Frezzotti, Garofalo, Kostrzewa, Pittler, Rossi, and Urbach numerically investigate an alternative to the Higgs mechanism in which the elementary particle masses are dynamically generated \cite{Frezzotti:2018zsy,Capitani:2018jtx}. Starting from an SU(3) gauge model,  an SU(2) doublet of non-abelian strongly interacting fermions is coupled to a complex scalar field doublet via a Yukawa and a Wilson-like term. In this model, the exact symmetry acting on all fields prevents power-divergent fermion mass terms. While the Yukawa and Wilson-like terms break the fermionic chiral invariance, chiral symmetry is restored at the critical Yukawa coupling up to effects of $O(\Lambda_{UV}^2)$. Assuming the scalar field has a double-well potential, the leftover chiral symmetry breaking at the cutoff scale polarizes the vacuum. This triggers spontaneous chiral symmetry breaking which generates a PCAC fermion mass. Such a dynamical fermion mass can be naturally ``small'' and also a natural hierarchy of 
fermion masses can emerge. Numerical simulations support the conjectured mechanisms \cite{Capitani:2018jtx}. In this set-up, the Higgs boson is a composite state in $WW+ZZ$ channel bound by new strongly interacting particles. Further, the inclusion of electro-weak interactions is explored in \cite{Frezzotti:2018zsy}.

  \subsection{Effects of a fundamental Higgs}

Maas and T\"orek point out that the physical spectrum must be gauge invariant. While in QCD this is guaranteed by confinement, the situation is different in the weak sector where the perturbative description is BRST-invariant, but gauge dependent. Experimental results match predictions due to the Fr\"ohlich-Morchio-Strocchi (FMS) mechanism \cite{Frohlich:1980gj,Frohlich:1981yi}. Because in the SM the weak gauge group matches the global custodial symmetry, physical and elementary spectrum are the same. For BSM models this is not guaranteed and could result in differences of physical and elementary spectrum. Investigating an SU(3) gauge theory with a fundamental Higgs field \cite{Maas:2018xxu}, Maas and T\"orek calculate the gauge invariant spectrum and compare it to predictions from gauge-invariant PT \cite{Maas:2017wzi} with the conclusion that standard PT fails to correctly predict the spectrum.

\section{Summary}
\label{Sec.summary}

Composite Higgs models explore the possibility that the Higgs boson is a bound state arising from a new strong sector. In general two scenarios are studied, the Higgs is a light scalar (dilaton-like) particle of the new strong dynamics or it arises similar to pions in QCD as a pseudo Nambu-Goldstone boson. In the end only the experiments will be able to tell us whether the Higgs is a composite particle. By performing nonperturbative simulations we can however guide experimentalists and model builders.

To meet current experimental constraints of a light 125 GeV Higgs boson but no observations of other, heavier resonances, it is conjectured that the new strong sector exhibits a large separation of scales which can arise from near-conformal dynamics. Simulations of near-conformal systems are more costly than QCD. A particular challenge is the identification of an IRFP at strong coupling using step-scaling techniques. While in principle a well defined method with little room for ambiguities, different groups arrive for the same system at different conclusions. Further work is required to understand the source of these discrepancies.

Results for the particle spectrum reveal a more consistent picture: simulations of near-confor\-mal systems exhibit a light iso-singlet scalar ($0^{++}$) with a mass similar to the pseudoscalar (pion) and well separated from the vector ($\rho$). The presence of a light scalar rules out chiral perturbation theory as low energy description and alternative effective field theories are explored. Moreover, models based on two representations or mass-split systems have revealed novel features, e.g.~chimera baryons combining constituents of both representations or a highly constrained particle spectrum exhibiting hyperscaling in a chirally broken system.

In order to better judge the viability of composite Higgs models as a description of the Higgs boson and the electro-weak sector in nature, it is important to use numerical lattice field theory simulations to extract phenomenologically testable quantities like the $S$ parameter or explore mechanisms to generate SM fermion masses. Furthermore, it would be tantalizing to establish relations between the Higgs sector and recently observed anomalies in the $B$-sector \cite{Amhis:2016xyh} (see also updates at \cite{HFLAV}) as discussed e.g.~by Marzocca \cite{Marzocca:2018wcf}.\vspace{-2mm}

\section*{Acknowledgments}\vspace{-2mm}
The author likes to thank T.~Appelquist, G.~Bergner, T.~DeGrand, V.~Drach, G.T.~Fleming, R.~Frezzotti, M.~Garofalo, M.~Golterman, D.C.~Hackett, A.~Hasenfratz, K.~Holland, W.I.~Jay, R.~Koniuk, J.~Kuti, J.W.~Lee, A.~Maas, Y.~Meurice, E.T.~Neil, D.~Nogradi, E.~Rinaldi, C.~Rebbi, and D.~Schaich for providing material for my talk as well as fruitful discussions. O.W.~acknowledges support by DOE grant DE-SC0010005.
\vspace{-2mm}

{\small
\bibliography{../General/BSM}
\bibliographystyle{apsrev4-1}
}



\end{document}